\documentclass[aps,preprint]{revtex4}
\usepackage[T1]{fontenc}
\usepackage{mathptmx}
\usepackage{graphicx}
\usepackage{subfigure}
\usepackage{float}
\usepackage{multirow}
\usepackage{amsmath,amsthm,mathrsfs,amsfonts,amssymb}
\usepackage{xcolor}
\usepackage[colorlinks,linkcolor=blue,citecolor=blue,urlcolor=blue]{hyperref}
\usepackage{cleveref}
\usepackage{anyfontsize}

\allowdisplaybreaks[4]
\DeclareMathAlphabet{\mathcal}{OMS}{cmsy}{m}{n}
\DeclareSymbolFont{Letters}{OML}{cmm}{m}{it}
\DeclareMathSymbol{\psi}{\mathalpha}{Letters}{32}
\DeclareMathSymbol{\Psi}{\mathalpha}{Letters}{9}
\crefname{equation}{Eq.}{Eqs.}
\crefname{figure}{Fig.}{Figs.}
\crefname{table}{Tab.}{Tabs.}
\crefname{section}{Sec.}{Secs.}

\DeclareMathOperator{\diag}{diag}
\newcommand{\beq}{\begin{equation}}
\newcommand{\eeq}{\end{equation}}
\newcommand{\ben}{\begin{align}}
\newcommand{\een}{\end{align}}
\newcommand{\bea}{\begin{aligned}}
\newcommand{\eea}{\end{aligned}}
\newcommand{\bes}{\begin{subequations}}
\newcommand{\ees}{\end{subequations}}
\newcommand{\bew}{\begin{widetext}}
\newcommand{\eew}{\end{widetext}}

\numberwithin{table}{section}

\begin{document}
\preprint{CTP-SCU/2021001}
\title{Holographic DC Conductivity for Backreacted NLED in Massive Gravity}
\author{Shihao Bi $^{a}$}
\email{bishihao@stu.scu.edu.cn}
\author{Jun Tao$^{a}$}
\email{taojun@scu.edu.cn}
\affiliation{$^{a}$Center for Theoretical Physics, College of Physics, Sichuan University, Chengdu, 610065, China}

\begin{abstract}
In this work a holographic model with the charge current dual to a general nonlinear electrodynamics (NLED) is discussed in the framework
of massive gravity. Massive graviton can breaks the diffeomorphism invariance in the bulk and generates momentum dissipation in the dual boundary theory.
The expression of DC conductivities in a finite magnetic field are obtained, with the backreaction of NLED field on the background geometry.
General transport properties in various limits are presented, and then we turn to the three of specific NLED models: the conventional Maxwell
electrodynamics, the Maxwell-Chern-Simons electrodynamics, and the Born-Infeld electrodynamics, to study the parameter-dependence of in-plane resistivity.
Two mechanisms leading to the Mott-insulating behaviors and negative magneto-resistivity are revealed at zero temperature, and the role 
played by the massive gravity coupling parameters are discussed.
\end{abstract}

\maketitle

\section{Introduction}
\label{sec:Intro}

The discovery of gauge/gravity duality makes it possible to deal with the strongly-coupled gauge theories on the boundary
from the classical gravitational theories in the higher dimensional bulk \cite{Susskind1997:PRD,Maldacena1997:IJTP,Witten1998:ATMP,Ammon2015:CUP,Eleftherios2011:LNP,Natsuume2016:LNP}.
And the well-known prediction on the ratio of the shear viscosity to the entropy density for $\mathcal{N}=4$ super Yang-Mills
(SYM) theory was found to be close to the experimental results of real quark-gluon plasma (QGP) \cite{Policastro2001:PRL,Buchel2004:PRL,Kovtun2005:PRL,Benincasa2006:JHEP}, 
making it more convincing to physical community. In recent years the idea of gauge/gravity duality has been applied to 
hydrodynamics \cite{DTSon2006:JHEP,DTSon2007:ARNPS,Kovtun2007:JPA,Rangamani2009:CQG},
quantum chromodynamics (QCD) \cite{Adams2001:PRD,Brodsky2004:PLB,Brodsky2005:PRL,Brodsky2009:PRL,Erlich2005:PRL,Rold20005:NPB,Zayakin2008:JHEP,Edelstein2009:AIP,Gursoy2011:QCD,Alfimov2015:JHEP}, nuclear physics \cite{Bergman2007:JHEP,Baldino2017:PRD}, 
and strongly-coupled condensed matter systems \cite{Hartnoll2008:PRL,Hartnoll2009:CQG,Herzog2009:JPA,Herzog2010:PRD,Mcgreevy2010:AHEP,Nishioka2010:JHEP,Cubrovic2009:Sci,Liu2011:PRD,Iqbal2012:NFL,Faulkner2011:JHEP,Cai2015:SCP}, etc, and new insights are brought into these physical branches. 

The establishment of quantum many-body theory \cite{AGD2012:QFT,Landau:vol9} in a fairly low energy scale compared with high energy physics 
is another magnificent and fascinating story apart from that of the standard model, which have profoundly extended and deepen our understanding of realistic matters. 
Many characterization methods, based on the linear response theory or quantum transport theory, have been developed in the scattering or transport experiments,
to reveal the electronic or lattice structures, surface topography, defects and disorder, and study the transport properties of the material samples.
In the framework of the energy band theory the materials are roughly divided into three categories: metals, semiconductors, and insulators, which differs in their
electrical conductivities. And it is known to all of us the classical electron motion in the presence of the magnetic field has been studied since 1879, 
which is the famous classical Hall effect \cite{Hall1879}. Its quantum version, known as the quantum Hall effect, was first observed in 1980 
by von Klitzing \cite{Klitzing1980:PRL}, has aroused a wide research enthusiasm in the last two decades and witnessed impressive
theoretical and experimental breakthroughs \cite{Qi2011:RMP,Hasan2010:RMP}.
The new quantum phases of matter, termed topological insulators, exhibit another kind of bulk-boundary correspondence between the gapped insulated
bulk and gapless metallic edge states on the boundary. Its exotic transport properties have also attracted lots of research interests, among which the 
conductivity behavior in the presence of magnetic field is of vital importance, for it provides a promising way on controlling the electrical properties
with the help of external fields. For normal metals the resistivity always increases with the magnetic field strength \cite{Wannier1972:PRB}, showing the positive magneto-resistivity. However, the experimental measurements \cite{Kim2013:PRL,Xiong2015:Sci,Li2016:NC,Zhang2016:NC,Zhao2016:SR} in topological materials, 
such as Dirac or Weyl semimetals, etc., demonstrated the presence of negative magneto-resistivity or crossover for positive to negative, which are
mainly attributed to chiral anomaly \cite{DTSon2013:PRB,Qi2013:CRP,Burkov2014:PRL,Burkov2015:PRB,Lu2017:FP}.
In addition, the negative magneto-resistivity behaviors are also found in holographic chiral anomalous systems \cite{Jimenez2014:PRD,Jimenez2015:JHEP,Landsteiner2015:JHEP,Sun2016:JHEP}.

Mott insulators, which are believed to be parent materials of cuprate high $T_c$ superconductors \cite{Wen2006:RMP}, 
is another example beyond the framework of energy band theory due to the strong Coulomb repulsive interaction.
The strong electron-electron interaction would prevent the available charge carriers to efficiently transport charges 
as if suffers the electronic traffic jam. The holographic construction of Mott insulators has been a heated topic
and much progress has been made \cite{Edalati2011:PRL,Edalati2011:PRD,Wu2012:JHEP,Wu2014:JHEP,Wu2015:JHEP,Wu2016:JHEP,Fujita2015:JHEP,Nishioka2010:JHEP,Kiritsis2015:JHEP}.
In Ref. \cite{Baggioli2016:JHEP}, the holographic model coupled with a particular NLED named iDBI was proposed to 
study the strong electron-electron interaction by introducing the self-interaction of NLED field, and the Mott-insulating
behaviors appear for large enough self-interaction strength.

In this paper, we construct a holographic model coupled with a general NLED field to investigate the magneto-transport
of the strongly-interacting system on the boundary from the perspective of gauge/gravity duality. The backreaction of NLED 
field on the bulk geometry is taken into consideration as in Ref. \cite{Cremonini2017:JHEP}. The massive gravity, with 
massive potentials associated with the graviton mass, is adopted to break the diffeomorphism invariance in the bulk 
producing momentum relaxation in the dual boundary theory \cite{Vegh2013}.

This paper is organized as follows. In \cref{sec:HS} we establish our holographic model of massive gravity with NLED field.
Then the DC conductivities with non-zero magnetic field are derived as the function of $\rho$, $h$, $T$, and related 
massive gravity coupling parameters in \cref{sec:DCC}, with various limit situation being discussed. Then in \cref{sec:NLED}
we present detailed investigation on the in-plane resistivity in the framework of conventional Maxwell electrodynamics,
the CP-violating Maxwell-Chern-Simons electrodynamics, and Born-Infeld electrodynamics. Finally, we make
our conclusion in \cref{sec:Conc}.
We will use the unit $\hslash=G=k_{B}=l=1$.

\section{Holographic Setup}
\label{sec:HS}

The 4-dimensional massive gravity \cite{Vegh2013} with a negative cosmological constant $\Lambda$ 
coupled to a nonlinear electromagnetic field $A_{\mu}$ we are considering is given by
\begin{equation}
S=\frac{1}{16\pi}\int\mathrm{d}^{4}x\sqrt{-g}\left[R-2\Lambda+m^{2}\sum_{j=1}^{4}c_{j}\mathcal{U}_{j}(g,f)+\mathcal{L}(s,p)\right] ,
\label{eq:action}
\end{equation}
where $\Lambda=-3/l^2$, $m$ is the mass parameter, $f$ is a fixed symmetric tensor called the reference metric, 
$c_{j}$ are coupling constants \footnote{To make the massive gravity theory self-consistent, all the $c_{j}$ will be set to be negative.}, 
and $\mathcal{U}_{j}(g,f)$ denotes the symmetric polynomials of the eigenvalue of the $4\times4$ matrix 
$\mathcal{K}_{\nu}^{\mu}=\sqrt{g^{\mu\lambda}f_{\lambda\nu}}$ given as
\begin{equation}
\begin{aligned}
\mathcal{U}_{1} & =[\mathcal{K}] ,\\
\mathcal{U}_{2} & =[\mathcal{K}]^{2}-\left[\mathcal{K}^{2}\right] ,\\
\mathcal{U}_{3} & =[\mathcal{K}]^{3}-3[\mathcal{K}]\left[\mathcal{K}^{2}\right]+2\left[\mathcal{K}^{3}\right] ,\\
\mathcal{U}_{4} & =[\mathcal{K}]^{4}-6[\mathcal{K}]^{2}\left[\mathcal{K}^{2}\right]+8[\mathcal{K}]\left[\mathcal{K}^{3}\right]+3\left[\mathcal{K}^{2}\right]^{2}-6\left[\mathcal{K}^{4}\right].
\end{aligned}
\end{equation}
The square root in $\mathcal{K}$ means $\left(\sqrt{\mathcal{K}}\right)_{\lambda}^{\mu}\left(\sqrt{\mathcal{K}}\right)_{\nu}^{\lambda}=\mathscr{\mathcal{K}}_{\nu}^{\mu}$
and $\left[\mathcal{K}\right]\equiv\mathcal{K}_{\mu}^{\mu}$. The NLED Lagrangian $\mathcal{L}(s,p)$ in \cref{eq:action} is constructed as the function of two independent
nontrivial scalar using the field strength tensor $F_{\mu\nu}=\partial_{\mu}A_{\nu}-\partial_{\nu}A_{\mu}$
\begin{subequations}\begin{align}
s = & -\frac{1}{4}F^{\mu\nu}F_{\mu\nu} , \\
p = & -\frac{1}{8}\varepsilon^{\mu\nu\rho\sigma}F_{\mu\nu}F_{\rho\sigma} .
\label{eq:scalar}
\end{align}\end{subequations}
Here $\epsilon^{abcd}\equiv-\left[a\;b\;c\;d\right]/\sqrt{-g}$ is a totally antisymmetric Lorentz tensor, 
and $\left[a\;b\;c\;d\right]$ is the permutation symbol. In the weak field limit we assume the NLED reduces 
to the Maxwell-Chern-Simons Lagrangian $\mathcal{L}\left(s,p\right)\approx s+\theta p$ with $\theta$ defined
as $\mathcal{L}^{\left(0,1\right)}\left(0,0\right)$ for later convenience.

Varying the action with respect to $g^{\mu\nu}$ and $A^{\mu}$ we obtain the equations of motion
\begin{subequations}\begin{align}
R_{\mu\nu}-\frac{1}{2}Rg_{\mu\nu}+\Lambda g_{\mu\nu}=& \frac{1}{2}T_{\mu\nu} , \label{eq:eom1} \\
\nabla_{\mu}G^{\mu\nu}=& 0 .
\label{eq:eom2}
\end{align}\end{subequations}
where the energy-momentum tensor is 
\begin{equation}
T_{\mu\nu}=g_{\mu\nu}\left(\mathcal{L}(s,p)-p\frac{\partial\mathcal{L}(s,p)}{\partial p}\right)+\frac{\partial\mathcal{L}(s,p)}{\partial s}F_{\mu}^{\lambda}F_{\nu\lambda}+m^{2}\chi_{\mu\nu},
\label{eq:EMT}
\end{equation}
with 
\begin{equation}\bea
\chi_{\mu\nu} = & c_{1}\left(\mathcal{U}_{1}g_{\mu\nu}-\mathcal{K}_{\mu\nu}\right) + c_{2}\left(\mathcal{U}_{2}g_{\mu\nu}-2\mathcal{U}_{1}\mathcal{K}_{\mu\nu}+2\mathcal{K}_{\mu\nu}^{2}\right) \\
& + c_{3}\left(\mathcal{U}_{3}g_{\mu\nu}-3\mathcal{U}_{2}\mathcal{K}_{\mu\nu}+6\mathcal{U}_{1}\mathcal{K}_{\mu\nu}^{2}-6\mathcal{K}_{\mu\nu}^{3}\right) \\
& + c_{4}\left(\mathcal{U}_{4}g_{\mu\nu}-4\mathcal{U}_{3}\mathcal{K}_{\mu\nu}+12\mathcal{U}_{2}\mathcal{K}_{\mu\nu}^{2}-24\mathcal{U}_{1}\mathcal{K}_{\mu\nu}^{3}+24\mathcal{K}_{\mu\nu}^{4}\right) .
\label{eq:MGT}
\eea\end{equation}
And we introduce
\begin{equation}
G^{\mu\nu}=-\frac{\partial\mathcal{L}(s,p)}{\partial F_{\mu\nu}}=\frac{\partial\mathcal{L}(s,p)}{\partial s}F^{\mu\nu}
+\frac{1}{2}\frac{\partial\mathcal{L}(s,p)}{\partial p}\varepsilon^{\mu\nu\rho\sigma}F_{\rho\sigma} .
\end{equation}
We are looking for a black brane solution with asymptotic AdS spacetime by taking the following ansatz \cite{Vegh2013,Cai2015:PRD} for the metric and the NLED field, and the reference metric:
\begin{subequations}\begin{align}
\mathrm{d}s^{2} =& -{}f(r)\mathrm{d}t^{2}+\frac{\mathrm{d}r^{2}}{f(r)}+r^{2}\left(\mathrm{d}x^{2}+\mathrm{d}y^{2}\right) , \label{eq:gmunu} \\
A =& A_{t}(r)\mathrm{d}t+\frac{h}{2}\left(x\mathrm{d}y-y\mathrm{d}x\right) , \label{eq:Amu} \\
f_{\mu\nu} =& \diag \left(0,0,\alpha^{2},\alpha^{2}\right). \label{eq:fmunu}
\end{align}\end{subequations} 
where $h$ is the magnetic field strength. 
From \cref{eq:gmunu,eq:Amu} we find the nontrivial scalars \cref{eq:scalar} are
\begin{subequations}\begin{align}
s = & \frac{1}{2}\left(A_{t}'(r)^{2}-\frac{h^{2}}{r^{4}}\right) , \\
p = & -\frac{hA_{t}'(r)}{r^{2}} .
\label{eq:scalar2}
\end{align}\end{subequations}
The equations of motion then are obtained as
\begin{subequations}\begin{align}
rf'(r)+f(r)-3r^{2} =& c_{1}\alpha m^{2} r+c_{2}\alpha^{2} m^{2} +\frac{r^{2}}{2}\left(A_{t}'(r)G^{rt}+\mathcal{L}(s,p)\right) , \label{eq:eom_tr} \\
rf''(r)+2f'(r)-6r  =& c_{1}\alpha m^{2} + r\left(\mathcal{L}(s,p)+hG^{xy}\right) , \label{eq:eom_xy} \\
\left[r^{2}G^{rt}\right]' =& 0 , \label{eq:eom_EM}
\end{align}\end{subequations} 
where the non-vanishing components $G^{\mu\nu}$ are
\begin{subequations}\begin{align}
G^{rt} =& \frac{\partial\mathcal{L}}{\partial p}\frac{h}{r^{2}}-\frac{\partial\mathcal{L}}{\partial s}A_{t}'(r), \label{eq:G^rt} \\
G^{xy} =& \frac{\partial\mathcal{L}}{\partial s}\frac{h}{r^{4}}+\frac{\partial\mathcal{L}}{\partial p}\frac{A_{t}'(r)}{r^{2}} . \label{eq:G^xy} 
\end{align}\end{subequations} 

\cref{eq:eom_EM} leads to $G^{tr}=\rho/r^2$ with $\rho$ being a constant. The event horizon $r_{h}$ is the root of $f(r)$, i.e., $f(r_{h})=0$, 
and the Hawking temperature of the black brane is given by 
\begin{equation}
T=\frac{f'(r_{h})}{4\pi} .
\label{eq:temp1}
\end{equation}
Then at $r=r_{h}$ \cref{eq:eom_tr} reduces to 
\begin{equation}
4\pi Tr_{h}-3r_{h}^{2}=c_{1}\alpha m^{2}r_{h}+c_{2}\alpha^{2}m^{2}+\frac{r_{h}^{2}}{2}\left(A_{t}'(r_{h})G_{h}^{rt}+\mathcal{L}(s_{h},p_{h})\right) ,
\label{eq:temp2}
\end{equation}
where 
\begin{subequations}\begin{align}
s_{h} = & \frac{1}{2}\left(A_{t}'(r_{h})^{2}-\frac{h^{2}}{r_{h}^{4}}\right) , \\
p_{h} = & -\frac{hA_{t}'(r_{h})}{r_{h}^{2}} , \\
G_{h}^{rt} = & \mathcal{L}^{\left(0,1\right)}\left(s_{h},p_{h}\right)\frac{h}{r_{h}^{2}}
-\mathcal{L}^{\left(1,0\right)}\left(s_{h},p_{h}\right) A_{t}'\left(r_{h}\right)
\end{align}\end{subequations}

\section{DC Conductivity}
\label{sec:DCC}

From the perspective of gauge/gravity duality, the black brane solution \cref{eq:gmunu,eq:Amu} in the bulk 
can describe an equilibrium state at finite temperature $T$ given by \cref{eq:temp2}. And the conserved current
$\mathcal{J}^{\mu}$ in the boundary theory is connected with the conjugate momentum of the NLED field in the bulk,
which allows us calculate the DC conductivity in the framework of linear response theory \cite{Donos2014:JHEP,Blake2014:PRL}.

\subsection{Derivation of DC Conductivity}
\label{subsec:der}

The following perturbations on the metric and the NLED field are applied to derive the DC conductivity:
\begin{subequations}\begin{align}
\delta g_{ti} = & r^{2}h_{ti}(r)  , \\
\delta g_{ri} = & r^{2}h_{ri}(r)  , \\
\delta A_{i} = & -E_{i}t+a_{i}(r) ,
\end{align}\end{subequations} 
where $i=x,y$. 
We first consider the $t$ component. The absence of $A_{t}(r)$ in the \cref{eq:scalar2} leads to the 
radially independent conjugate momentum $\partial_{r}\Pi^{i}=0$, with 
\begin{equation}
\Pi^{t}=\frac{\partial\mathcal{L}\left(s,p\right)}{\partial\left(A_{t}^{\prime}\left(r\right)\right)}. 
\label{eq:Pit}
\end{equation}
Then the expectation value of $\mathcal{J}^{t}$ in the dual boundary field theory is given by 
\begin{equation}
\left\langle \mathcal{J}^{t}\right\rangle =\Pi^{t}.
\end{equation}
In the linear level we have $\left\langle \mathcal{J}^{t}\right\rangle=\rho$, which indicate that $\rho$
can be interpreted as the charge density in the dual field theory. At the event horizon $r=r_{h}$ the 
charge density $\rho$ is given by 
\begin{equation}
\rho=\mathcal{L}^{\left(1,0\right)}\left(s_{h},p_{h}\right)r_{h}^{2}A_{t}^{\prime}\left(r_{h}\right)-\mathcal{L}^{\left(0,1\right)}\left(s_{h},p_{h}\right)h.
\label{eq:rho}
\end{equation}

Then we consider the planar components. The NLED is explicitly independent of $a_{i}(r)$, making the 
conjugate momentum of the field $a_{i}(r)$
\begin{equation}
\Pi^{i}=\frac{\partial\mathcal{L}\left(s,p\right)}{\partial\left(a_{i}^{\prime}(r)\right)}
=\frac{\partial\mathcal{L}\left(s,p\right)}{\partial\left(\partial_{r}A_{i}\right)}=\sqrt{-g}G^{ir},
\end{equation}
being radially independent as well. And the charge currents in the dual field theory are given by $\left\langle \mathcal{J}^{i}\right\rangle =\Pi^{i}$,
and can be expressed with the perturbed metric and field components $h_{ti}$, $h_{ri}$, $a'_{i}$ and $E_{i}$:
\begin{subequations}\begin{align}
\left\langle \mathcal{J}^{x}\right\rangle &=-\mathcal{L}^{\left(1,0\right)}\left(s,p\right)\left[f\left(r\right)a_{x}^{\prime}\left(r\right)+hf\left(r\right)h_{ry}\left(r\right)+r^{2}A_{t}^{\prime}\left(r\right)h_{tx}\left(r\right)\right]-\mathcal{L}^{\left(0,1\right)}\left(s,p\right)E_{y} , \label{eq:Jx} \\
\left\langle \mathcal{J}^{y}\right\rangle &=-\mathcal{L}^{\left(1,0\right)}\left(s,p\right)\left[f\left(r\right)a_{y}^{\prime}\left(r\right)-hf\left(r\right)h_{rx}\left(r\right)+r^{2}A_{t}^{\prime}\left(r\right)h_{ty}\left(r\right)\right]+\mathcal{L}^{\left(0,1\right)}\left(s,p\right)E_{x} . \label{eq:Jy}
\end{align}\end{subequations} 
The perturbed metric components $h_{ti}$ and $h_{ri}$ are coupled to $E_{i}$ in gravitational field equation and can be reduced. 
Thus we take the first order of the perturbed Einstein's equations of $tx$ and $ty$ components and we get:
\begin{subequations}\begin{align}
\left(\frac{h^{2}}{r^{2}}-\frac{\alpha m^{2} c_{1}r}{\mathcal{L}^{\left(1,0\right)}(s,p)}\right)h_{tx}(r)-hA_{t}'(r)f(r)h_{ry}(r)=&A_{t}'(r)f(r)a'_{x}(r)-\frac{E_{y}h}{r^{2}} , \label{eq:tx_p}\\
\left(\frac{h^{2}}{r^{2}}-\frac{\alpha m^{2} c_{1}r}{\mathcal{L}^{\left(1,0\right)}(s,p)}\right)h_{ty}(r)+hA_{t}'(r)f(r)h_{rx}(r)=&A_{t}'(r)f(r)a'_{y}(r)+\frac{E_{x}h}{r^{2}} . \label{eq:ty_p}
\end{align}\end{subequations} 
Next by using the regularity constraints of the metric and field near the event horizon \cite{Blake2014:PRL}:
\begin{equation}\begin{aligned}
f\left(r\right) & =4\pi T\left(r-r_{h}\right)+\cdots,\\
A_{t}\left(r\right) & =A_{t}^{\prime}\left(r_{h}\right)\left(r-r_{h}\right)+\cdots,\\
a_{i}\left(r\right) & =-\frac{E_{i}}{4\pi T}\ln\left(r-r_{h}\right)+\cdots,\\
h_{ri}\left(r\right) & =\frac{h_{ti}\left(r\right)}{f\left(r\right)}+\cdots,
\label{eq:reg}
\end{aligned}\end{equation}
the \cref{eq:tx_p,eq:ty_p} at the event horizon $r=r_{h}$ are
\begin{subequations}\begin{align}
hA_{t}'(r_{h})h_{ty}(r_{h})-\left(\frac{h^{2}}{r_{h}^{2}}-\frac{\alpha m^{2} c_{1}r_{h}}{\mathcal{L}^{\left(1,0\right)}(s_{h},p_{h})}\right)h_{tx}(r_{h})=A_{t}'(r_{h})E{}_{x}+\frac{h}{r_{h}^{2}}E_{y} , \label{eq:tx_p2}\\
\left(\frac{h^{2}}{r_{h}^{2}}-\frac{\alpha m^{2} c_{1}r_{h}}{\mathcal{L}^{\left(1,0\right)}(s_{h},p_{h})}\right)h_{ty}(r_{h})+hA_{t}'(r_{h})h_{tx}(r_{h})=\frac{h}{r_{h}^{2}}E_{x}-A_{t}'(r_{h})E{}_{y} . \label{eq:ty_p2}
\end{align}\end{subequations} 
Solving \cref{eq:tx_p2,eq:ty_p2} for $h_{ti}(r_{h})$ in terms of $E_{i}$ and inserting into \cref{eq:Jx,eq:Jy}, one can evaluate the current $\left\langle \mathcal{J}^{i}\right\rangle$ to the electric fields $E_{i}$ at the event horizon $r=r_{h}$ via $\left\langle \mathcal{J}^{i}\right\rangle = \sigma_{ij} E_{j}$, where the DC conductivities are given by
\begin{subequations}\begin{align}
\sigma_{xx} =& \sigma_{yy}= \dfrac{\alpha m^{2} c_{1}r_{h}\left(\dfrac{\alpha m^{2} c_{1}r_{h}}{\mathcal{L}^{\left(1,0\right)}(s_{h},p_{h})}-\dfrac{h^{2}}{r_{h}^{2}}+r_{h}^{2}A'_{t}{}^{2}(r_{h})\right)}{\left(\dfrac{\alpha m^{2} c_{1}r_{h}}{\mathcal{L}^{\left(1,0\right)}(s_{h},p_{h})}-\dfrac{h^{2}}{r_{h}^{2}}\right)^{2}+h^{2}A'_{t}{}^{2}(r_{h})} , \label{eq:cond_xx}\\
\sigma_{xy} =&-\sigma_{yx}= \dfrac{\mathcal{L}^{\left(1,0\right)}(s_{h},p_{h})r_{h}^{2}A'_{t}(r_{h})}{h}\left(1-\dfrac{\left(\dfrac{\alpha m^{2} c_{1}r_{h}}{\mathcal{L}^{\left(1,0\right)}(s_{h},p_{h})}\right)^{2}}{\left(\dfrac{\alpha m^{2} c_{1}r_{h}}{\mathcal{L}^{\left(1,0\right)}(s_{h},p_{h})}-\dfrac{h^{2}}{r_{h}^{2}}\right)^{2}+h^{2}A'_{t}{}^{2}(r_{h})}\right)-\mathcal{L}^{\left(0,1\right)}(s_{h},p_{h}) . \label{eq:cond_xy}
\end{align}\end{subequations} 
The resistivity matrix is the inverse of the conductivity matrix:
\begin{equation}
R_{xx}=R_{yy}=\frac{\sigma_{xx}}{\sigma_{xx}^{2}+\sigma_{xy}^{2}}
\quad\text{and}\quad
R_{xy}=-R_{yx}=\frac{\sigma_{xy}}{\sigma_{xx}^{2}+\sigma_{xy}^{2}} . 
\end{equation}

The event horizon radius $r_{h}$ is the solution of \cref{eq:temp2} and will be affected by the temperature $T$, 
the charge density $\rho$, the magnetic field $h$ and the parameters $c_{1,2}$, $m$ and $\alpha$. 
For a given Lagrangian $\mathcal{L}(s,p)$ one first solve $A_{t}'(r_{h})$ from \cref{eq:rho}, and then plug it into
\cref{eq:temp2} to solve $r_h$, which will be brought into \cref{eq:cond_xx,eq:cond_xy} to obtain the DC conductivity
as a complicated function of the parameters mentioned above. 

\subsection{Various Limits}
\label{subsec:vl}

The concrete examples of NLED will be discussed in \cref{sec:NLED} and the properties of DC resistivity are discussed in detail.
Before focus on the specific NLED model we first consider some general properties of the DC conductivity in various limit.

\subsubsection{Massless and Massive Limits}
\label{subsubsec:MML}

The massless limit corresponds to $m\to0$ and the system will restore the Lorentz invariance. 
It has been shown that in a Lorentz invariant theory that the DC conductivities in the presence 
of a magnetic field were $\sigma_{xx}=0$ and $\sigma_{xy}=\rho/h$ \cite{Hartnoll2007:PRD}. 
For comparison, in the massless limit the DC conductivities in \cref{eq:cond_xx,eq:cond_xy} become 
\begin{equation}
\sigma_{xx}=\frac{\alpha m^{2}\left|c_{1}\right|r_{h}^{3}}{h^{2}}+\mathcal{O}(m^{4})
\quad\text{and}\quad
\sigma_{xy}=\frac{\rho}{h}+\mathcal{O}(m^{4}) .
\end{equation}
And in the massive limit $m^{-1}\to+\infty$, the DC conductivities have the asymptotic behaviors
\begin{equation}
\sigma_{xx}=\mathcal{L}^{\left(1,0\right)}(s_{h},p_{h})+\mathcal{O}(m^{-2})
\quad\text{and}\quad
\sigma_{xy}=-\mathcal{L}^{\left(0,1\right)}(s_{h},p_{h})+\mathcal{O}(m^{-2}) .
\end{equation}
which agree with the calculation by treating the NLED field as a probe one \cite{Guo2017:PRD}, 
for the geometry is dominated by the massive terms.

\subsubsection{Zero Field and Charge Density Limits}
\label{subsubsec:ZFCDL}

In the zero field limit $h=0$ case, the DC conductivities become
\begin{equation}
\sigma_{xx}=\mathcal{L}^{\left(1,0\right)}\left(\frac{A_{t}^{\prime2}\left(r_{h}\right)}{2
},0\right)-\frac{\rho^{2}}{\alpha m^{2}c_{1}r_{h}^{3}}
\quad\text{and}\quad
\sigma_{xy}=-\mathcal{L}^{\left(0,1\right)}\left(\frac{A_{t}^{\prime2}\left(r_{h}\right)}{2
},0\right) .
\end{equation}
where $A_{t}^{\prime}\left(r_{h}\right)$ is obtained by solving
\begin{equation}
\rho=\mathcal{L}^{\left(1,0\right)}\left(\frac{A_{t}^{\prime2}\left(r_{h}\right)}{2},0\right)r_{h}^{2}A_{t}^{\prime}\left(r_{h}\right).
\end{equation}

At zero charge density $\rho=0$, the DC conductivities become
\begin{equation}
\sigma_{xx}^{-1}=\frac{1}{\mathcal{L}^{\left(1,0\right)}\left(-\dfrac{h^{2}}{2r_{h}^{4}},0\right)}-\dfrac{h^{2}}{\alpha m^{2}c_{1}r_{h}^{3}}
\quad\text{and}\quad
\sigma_{xy}=-\mathcal{L}^{\left(0,1\right)}\left(-\frac{h^{2}}{2r_{h}^{4}},0\right) .
\end{equation}
We find that the DC conductivities are in general non-zero and can be interpreted as
incoherent contributions \cite{Davison2015:JHEP}, known as the charge
conjugation symmetric contribution $\sigma_{\text{ccs}}$. There is another
contribution from explicit charge density relaxed by some momentum
dissipation, $\sigma_{\text{diss}}$, depending on the charge density $\rho$. The
results show that, for a general NLED model, the DC conductivities usually
depend on $\sigma_{\text{diss}}$ and $\sigma_{\text{ccs}}$ in a nontrivial way.

\subsubsection{High Temperature Limit}
\label{subsubsec:HTL}

\begin{figure}[H]
\centering
\subfigure{\begin{minipage}[t]{0.45\textwidth}
\centering
\includegraphics[scale=0.5]{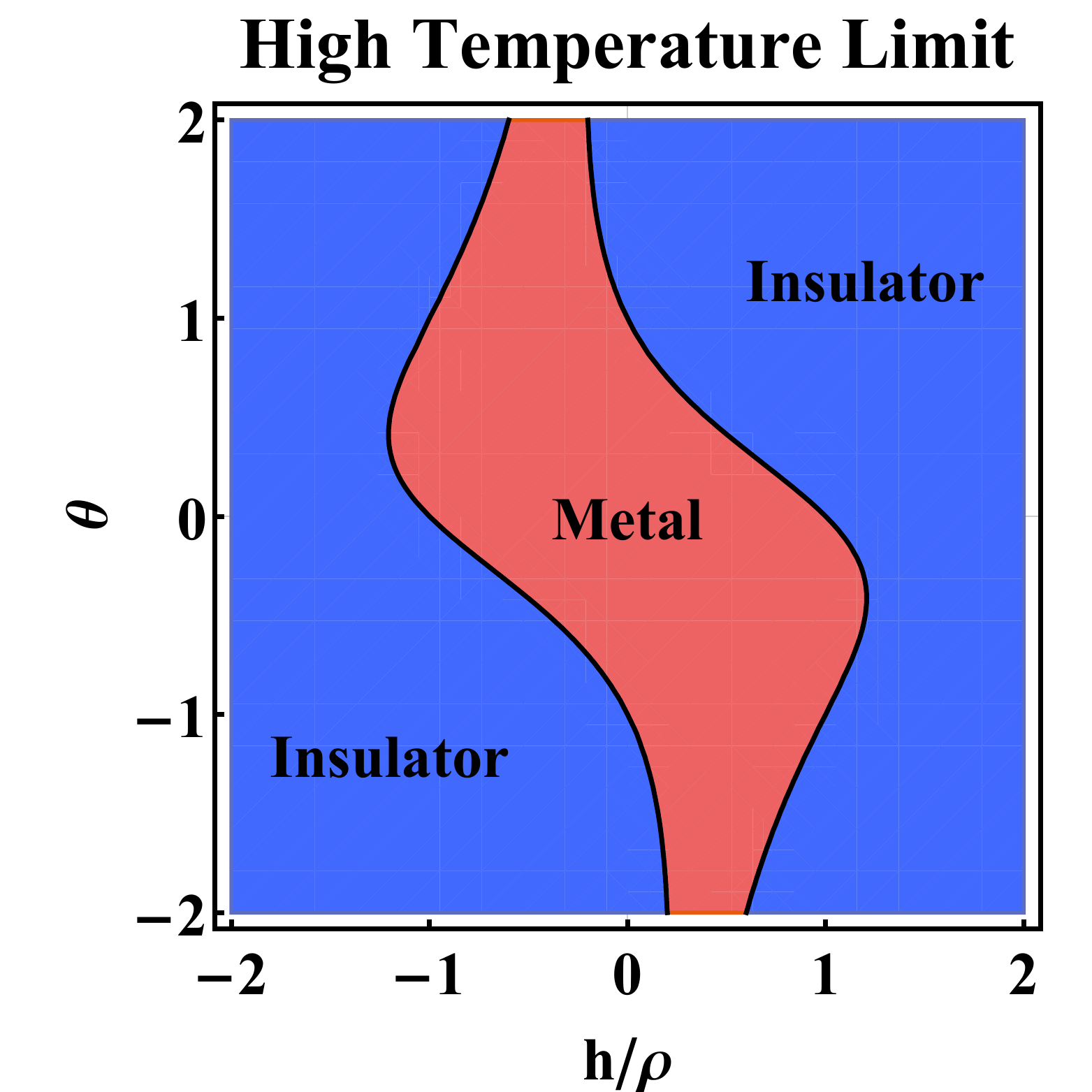}
\end{minipage}
} 
\subfigure{\begin{minipage}[t]{0.45\textwidth}
\centering
\includegraphics[scale=0.5]{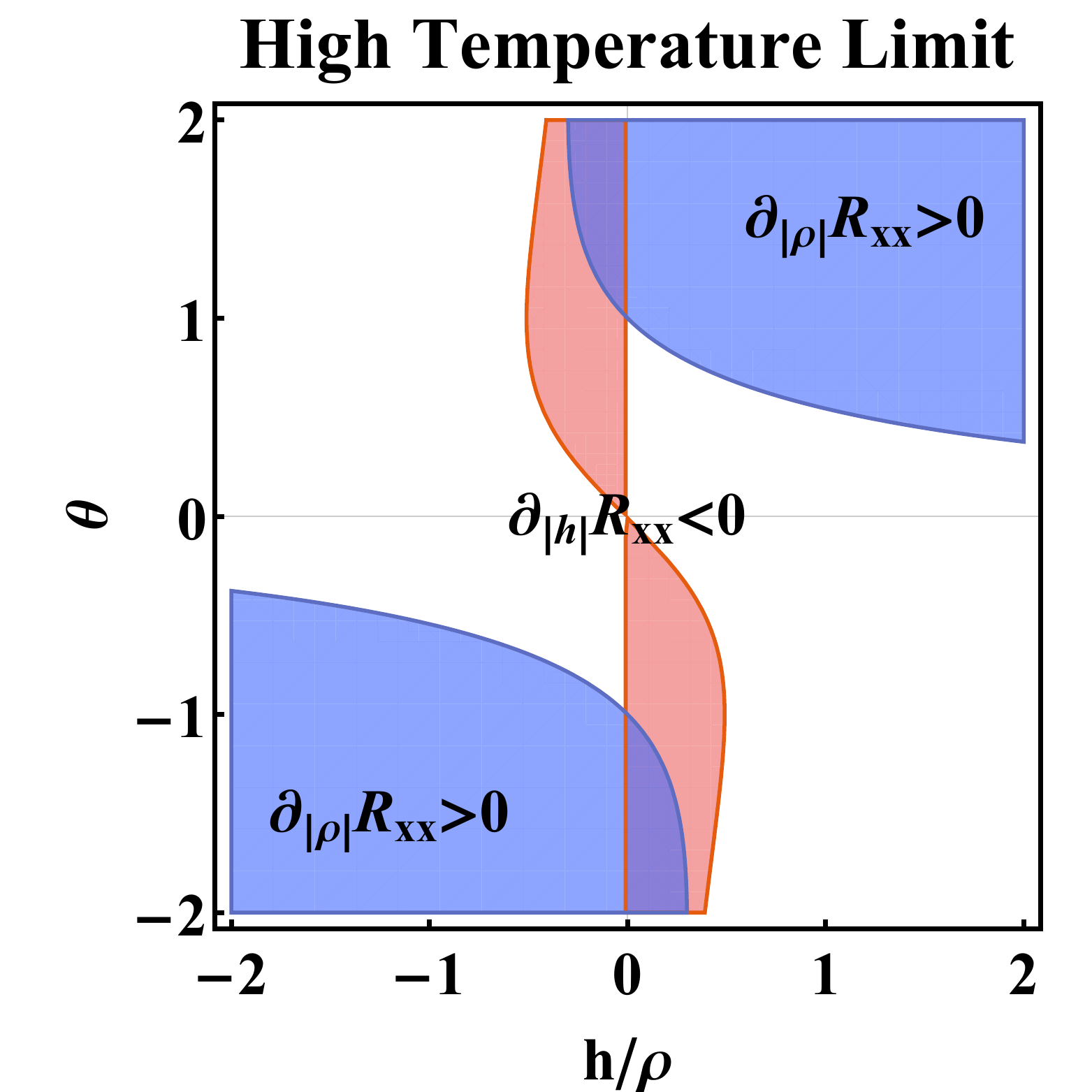}
\end{minipage}
}
\caption{Left panel: The metal-insulator phase diagram in the high temperature limit. 
The red region has positive temperature derivative of $R_{xx}$ and thus describes a metal, 
while the blue region has negative temperature derivative, and hence an insulator. The black solid
lines are the phase boundaries. 
Right panel: The Mott-insulating region (blue) and the negative magneto-resistivity (red) in the high
temperature limit}
\label{fig:MM}
\end{figure}

Finally, we consider the high temperature limit $T\gg\max\left\{\sqrt{h},\sqrt{\rho},|c_1|\alpha m^2,\sqrt{|c_2|}\alpha m\right\}$. 
In this limit, \cref{eq:temp2} gives $T\approx\dfrac{3}{4\pi}r_{h}$.
The longitudinal resistivity then reduces to
\begin{equation}
R_{xx}=\frac{1}{1+\theta^{2}}\left\{ 1+\frac{27}{64\pi^{3}\alpha\left|c_{1}\right|T^{3}}\left[h^{2}\left(1+\theta^{2}\right)+
2h\theta\rho-\frac{1-\theta^{2}}{1+\theta^{2}}\rho^{2}\right]\right\} +\mathcal{O}\left(T^{-6}\right) . \label{eq:Rxx}
\end{equation}
The nonlinear effect will be suppressed by the temperature and we only keep the leading order. Usually the metal and insulator
have different temperature dependence. For the metallic materials the phonon scattering enlarges the resistivity, while the
thermal excitation of carriers in insulating materials can promotes the conductivity. And the metal-insulator transition
can occur when the coefficient of $1/T^2$ changes the sign, which is shown in \cref{fig:MM} in the $h/\rho-\theta$ parameter
space. The $\theta$ term can break the $\left(\rho,h\right)  \to \left(\rho,-h\right)$ or $\left(\rho,h\right) \to \left(-\rho,h\right)$ 
symmetries for $\sigma_{ij}$ or $R_{ij}$, but $\sigma_{ij}$ or $R_{ij}$ are invariant under $\left(\rho,h\right) \to \left(-\rho,-h\right)$.
And the phases are central symmetric in the parameter plane.

The Mott-insulating and magneto-resistance behaviors are also presented in \cref{fig:MM}. The Mott-insulating region
is where $\partial_{|\rho|}R_{xx}>0$ in the parameter plane. The strong electron-electron interaction prevents
the charge carriers from transporting. From \cref{eq:Rxx} we see that as long as $\theta^2>1$ the Mott-insulating
emerges in the absence of the magnetic field.
The magneto-resistance is defined as 
\begin{equation}
MR_{xx} = \frac{R_{xx}\left(h\right)-R_{xx}\left(0\right)}{R_{xx}\left(0\right)},
\end{equation}
and we see that in the \cref{fig:MM} the negative magneto-resistivity can only occur with non-zero $\theta$.
For $\theta=0$, \cref{eq:Rxx} reduces to 
\begin{equation}
R_{xx} = 1+\frac{27\left(h^{2}-\rho^{2}\right)}{64\pi^{3}\alpha\left|c_{1}\right|T^{3}}+\mathcal{O}\left(T^{-6}\right) ,
\label{eq:Rxx_2}
\end{equation}
which gives metallic behavior for $h|/\rho|<1$ and insulating behavior for $|/\rho|>1$. And one always has
$\partial_{|\rho|}R_{xx}<0$ and $\partial_{|h|}R_{xx}>0$, indicating the absence of Mott-insulating behavior
and negative magneto-resistivity.

\section{Various NLED Models}
\label{sec:NLED}

In this section, we will use \cref{eq:temp2,eq:rho,eq:cond_xx,eq:cond_xx} to study the dependence of the in-plane
resistivity $R_{xx}$ on the temperature $T$, the charge density $\rho$ and the magnetic field $h$ in some specific 
NLED models. For convenience we rescale the $c_{i}$'s as $c_{1}\sim\alpha m^{2}c_{1}$ and $c_{2}\sim\alpha^{2}m^{2}c_{2}$.
The conventional Maxwell electrodynamics is firstly presented, with a detail discussion on the $R_{xx}$ and $R_{xy}$'s 
dependence on involved massive gravity coupling parameters $c_1$ and $c_2$, the charge density $\rho$, the magnetic
field $h$ and the temperature $T$. Then the Chern-Simons $\theta$ term as an extension is introduced to investigate
the CP-violating effect. Finally, we discuss the Born-Infeld electrodynamics and the influence of non-linear effect 
on the DC resistivity.
The high temperature behaviors have been discussed in \cref{subsubsec:HTL}, so we will mainly focus on the
behavior of $R_{xx}$ around $T=0$ in this section.

\subsection{Maxwell Electrodynamics}
\label{subsec:ME}

\begin{figure}[H]
\centering
\subfigure{\begin{minipage}[t]{0.45\textwidth}
\centering
\includegraphics[scale=0.3]{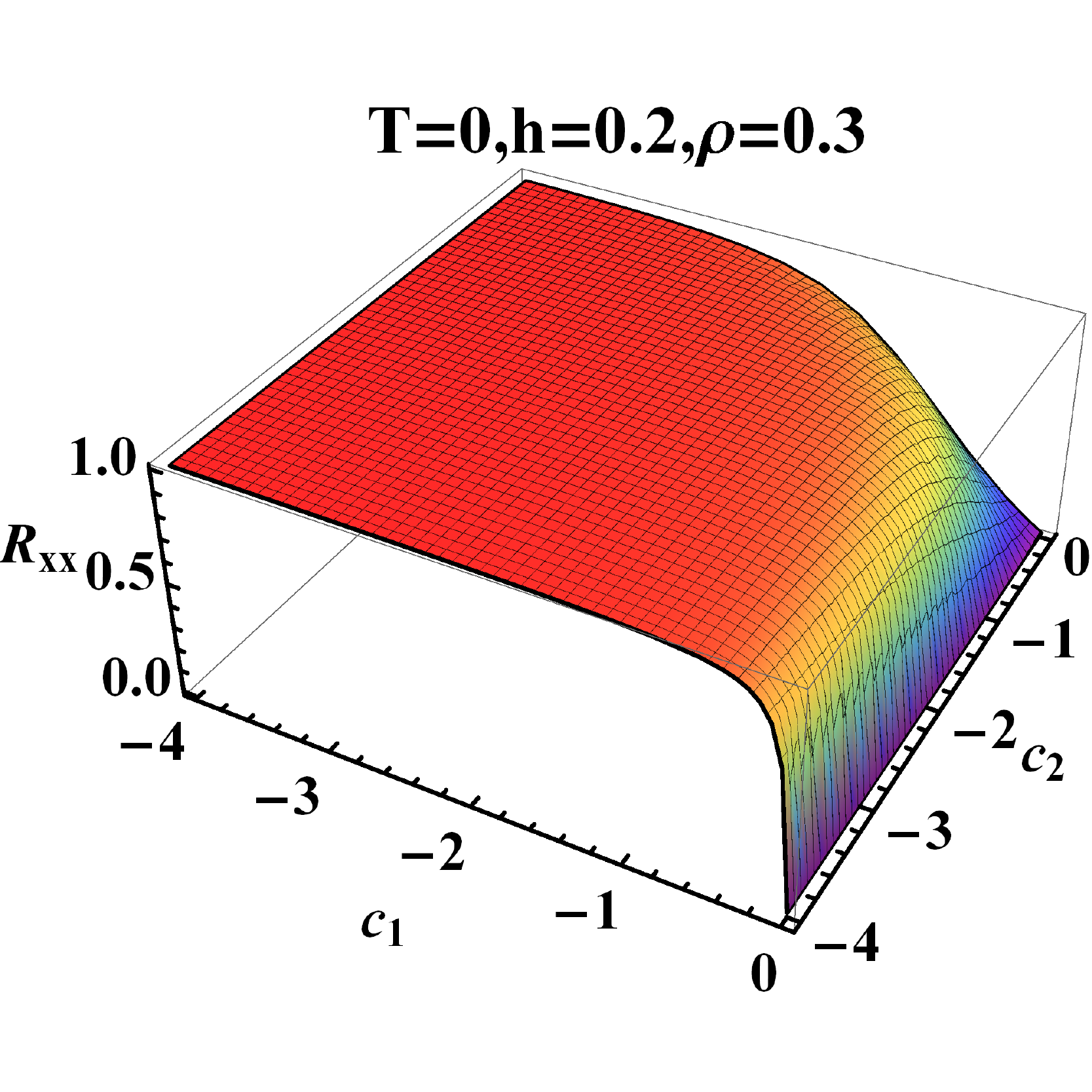}
\end{minipage}
} 
\subfigure{\begin{minipage}[t]{0.45\textwidth}
\centering
\includegraphics[scale=0.3]{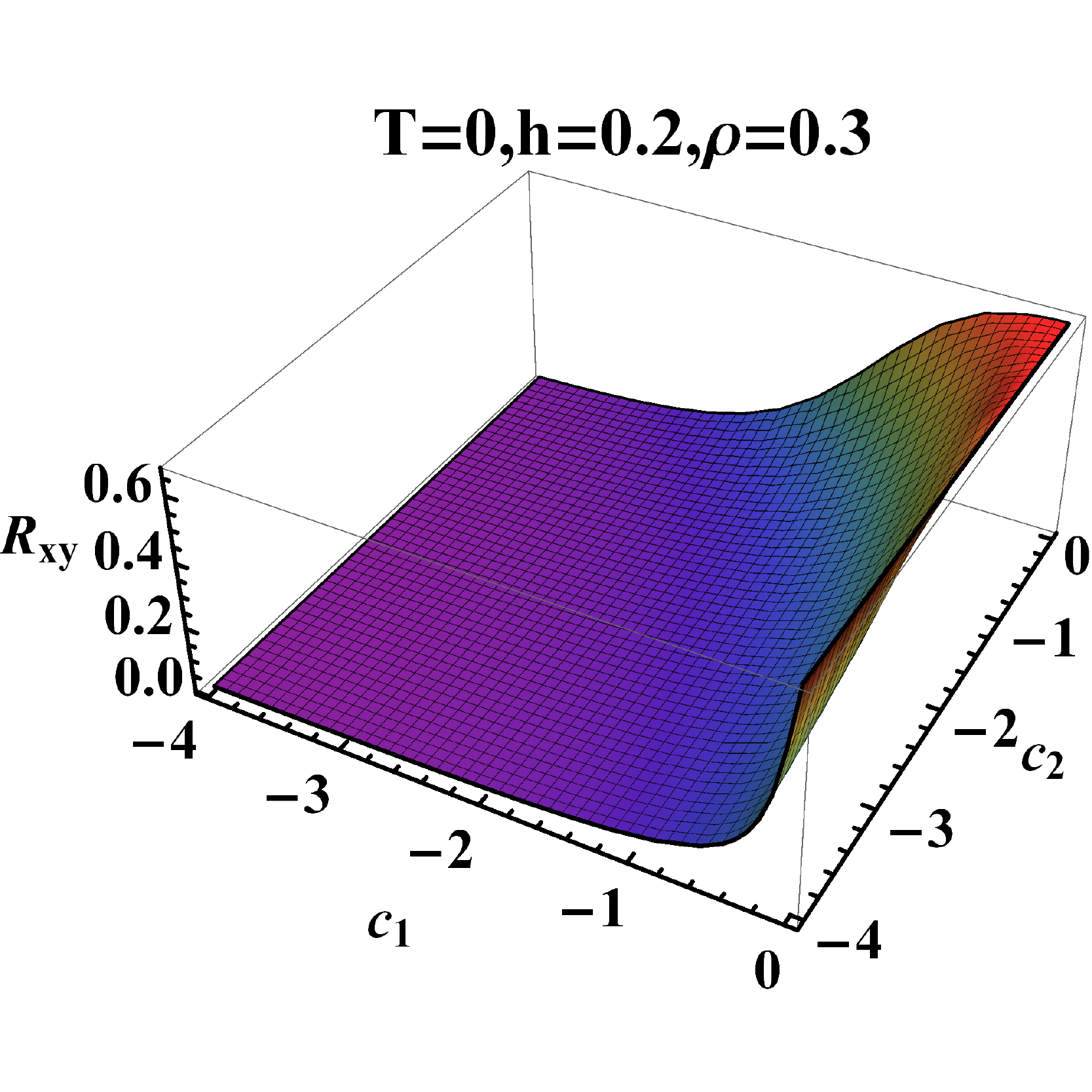}
\end{minipage}
}
\caption{The influence of $c_1$ and $c_2$ on $R_{xx}$ and $R_{xy}$. We take $T=0$ and $\rho=0.3$,$h=0.2$ as an example.}
\label{fig:M_R_0}
\end{figure}

\begin{figure}[H]
\centering
\includegraphics[width=14cm]{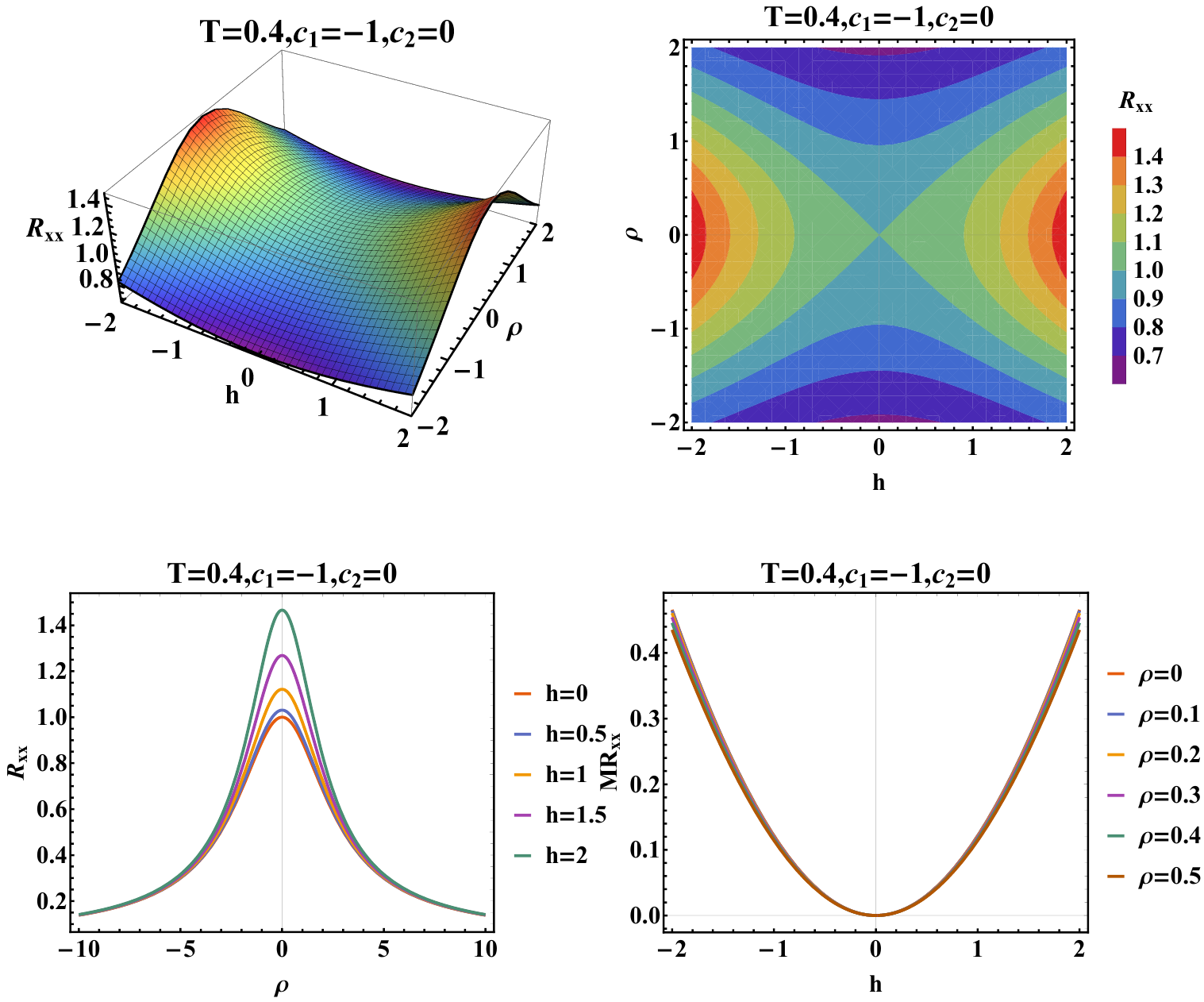}
\caption{Upper left: $R_{xx}$ as a function of $\rho$ and $h$. Upper right: Contour plot of $R_{xx}$.
Lower left: $R_{xx}$ vs $\rho$ with $h=0,0.5,1,1.5,2$. Lower right: $MR_{xx}$ vs $h$ with $\rho=0,0.1,0.2,0.3,0.4,0.5$.
We set $T=0.4$, $c_1=-1$, and $c_2=0$.}
\label{fig:M_Rxx_2}
\end{figure}

\begin{figure}[H]
\centering
\includegraphics[width=14cm]{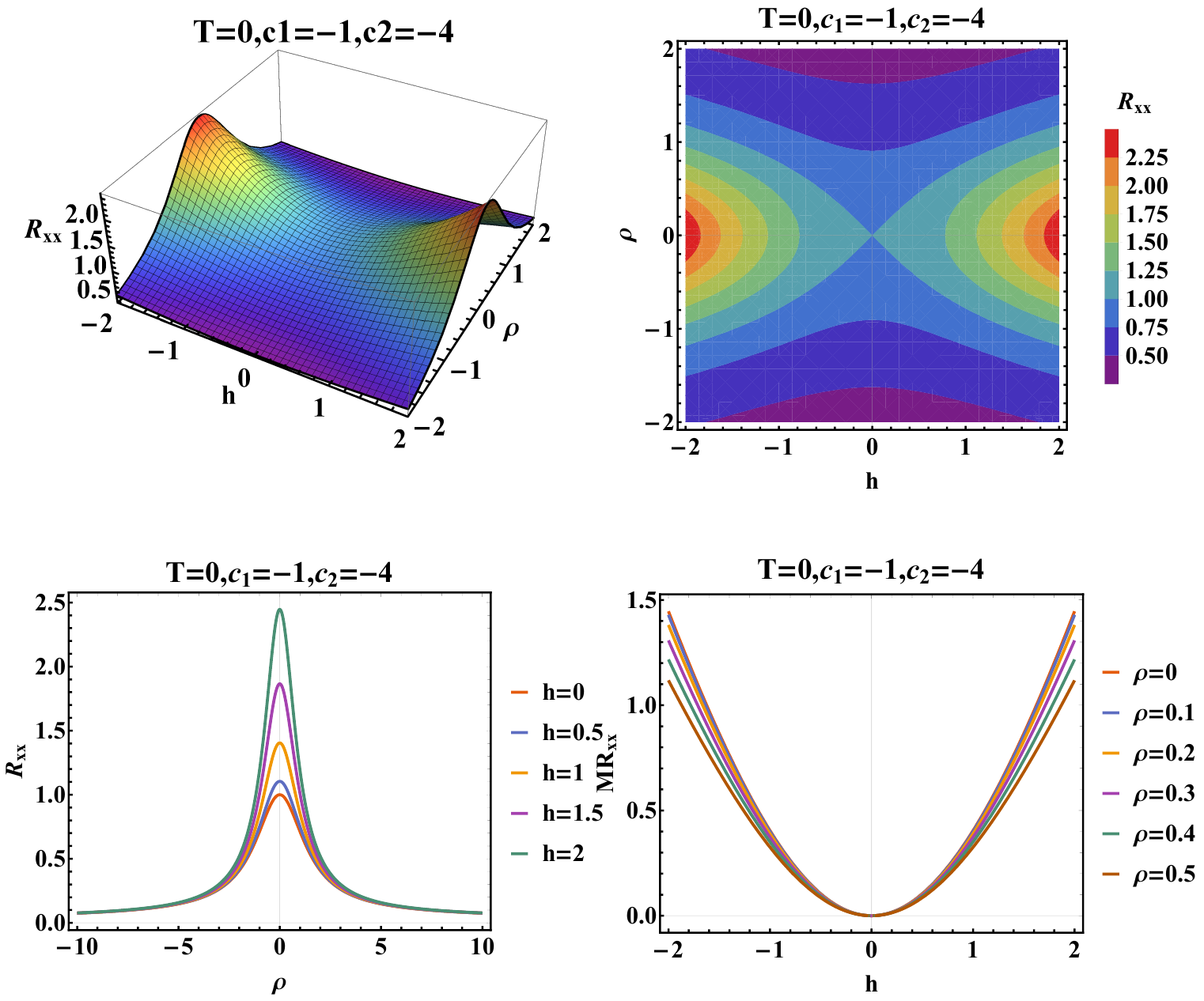}
\caption{Upper left: $R_{xx}$ as a function of $\rho$ and $h$. Upper right: Contour plot of $R_{xx}$.
Lower left: $R_{xx}$ vs $\rho$ with $h=0,0.5,1,1.5,2$. Lower right: $MR_{xx}$ vs $h$ with $\rho=0,0.1,0.2,0.3,0.4,0.5$.
We set $T=0$, $c_1=-1$, and $c_2=-4$.}
\label{fig:M_Rxx_3}
\end{figure}

We first consider the Maxwell electrodynamics, in which $\mathcal{L}(s,p)=s$. From \cref{eq:rho} we can solve out that
\begin{equation}
A_{t}^{\prime}\left(r_{h}\right)=\frac{\rho}{r_{h}^{2}} , 
\end{equation}
and bring it into \cref{eq:temp2} we get the equation
\begin{equation}
4\pi Tr_{h}-3r_{h}^{2}-c_{1} r_{h}-c_{2} +\frac{\rho^{2}+h^{2}}{4r_{h}^{2}} = 0 . 
\end{equation}
It is notice that $c_1$ offers an effective temperature correction, and make the solution complicated even at $T=0$.
Although the $c_{2}$ is a constant playing a similar role as the momentum dissipation strength in Ref. \cite{Peng2018:EPJC},
the effect on the resistivity is quite different. The effect of $c_1$ and $c_2$ on $R_{xx}$ and $R_{xy}$ at zero temperature
is presented in \cref{fig:M_R_0}. For $c_1=0$ the $R_{xx}=0$ and $R_{xy}$ is constant, and is independent of $c_2$. 
For non-zero $c_1$, the $R_{xx}$ increases and saturates, while the $R_{xy}$ decreases to zero as $c_1$ becomes more negative. 
And a larger $|c_2|$ makes the surface steeper.
\begin{figure}[H]
\centering
\includegraphics[width=14cm]{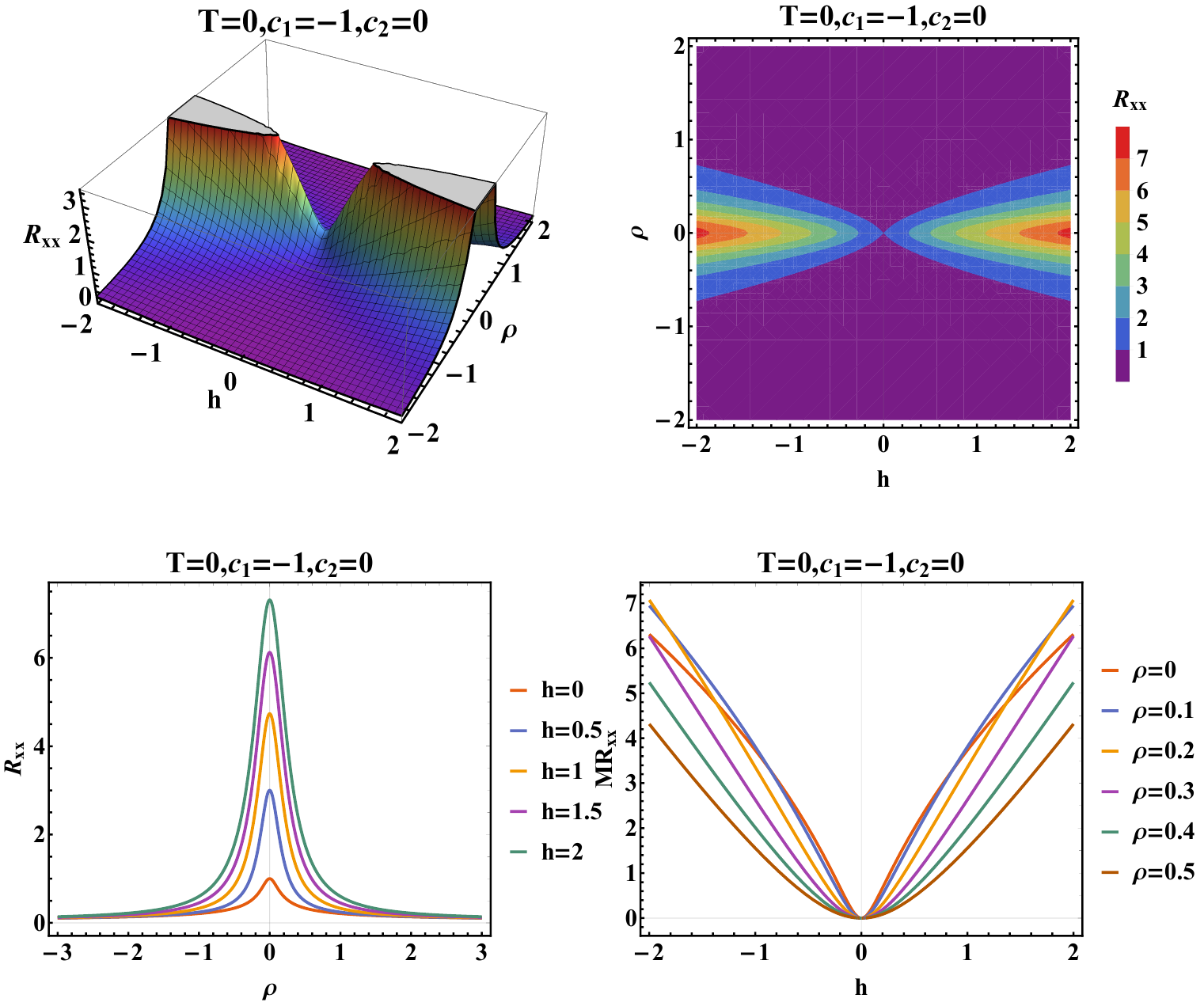}
\caption{Upper left: $R_{xx}$ as a function of $\rho$ and $h$. Upper right: Contour plot of $R_{xx}$.
Lower left: $R_{xx}$ vs $\rho$ with $h=0,0.5,1,1.5,2$. Lower right: $MR_{xx}$ vs $h$ with $\rho=0,0.1,0.2,0.3,0.4,0.5$.
We set $T=0$, $c_1=-1$, and $c_2=0$.}
\label{fig:M_Rxx_1}
\end{figure}
In the following \cref{fig:M_Rxx_1,fig:M_Rxx_2,fig:M_Rxx_3} we separately investigate
the resistivities as functions of $\rho$ and $h$ under different $T$ and $c_2$.
Because $c_1$ acts as an effective temperature modification, we set $c_1=-1$.

First, we consider zero temperature $T=0$, and $c_2=0$. No Mott-insulating behaviors and positive magneto-resistivity 
is observed in \cref{fig:M_Rxx_1}. In addition, non-zero charge density can suppress the resistivity as more charge 
carriers are introduced.
In \cref{fig:M_Rxx_3} the effect of $c_2$ at zero temperature is presented. The figures show similar 
feature as those in \cref{fig:M_Rxx_2}. The larger $c_2$ also suppress the resistivity, while the influence
is not so obvious as the temperature. All the pictures presented here do not exhibit the Mott-insulating behaviors
and negative magneto-resistivities.
Then the finite temperature situation with $T=0.4$ and $c_2=0$ is studied in \cref{fig:M_Rxx_2}, 
where the saddle surface is relatively flatter than that in \cref{fig:M_Rxx_1}. And the magneto-resistivity
is greatly suppressed by the temperature and the differences induced by charge density $\rho$ is smoothen out.

\begin{figure}[H]
\centering
\includegraphics[width=14cm]{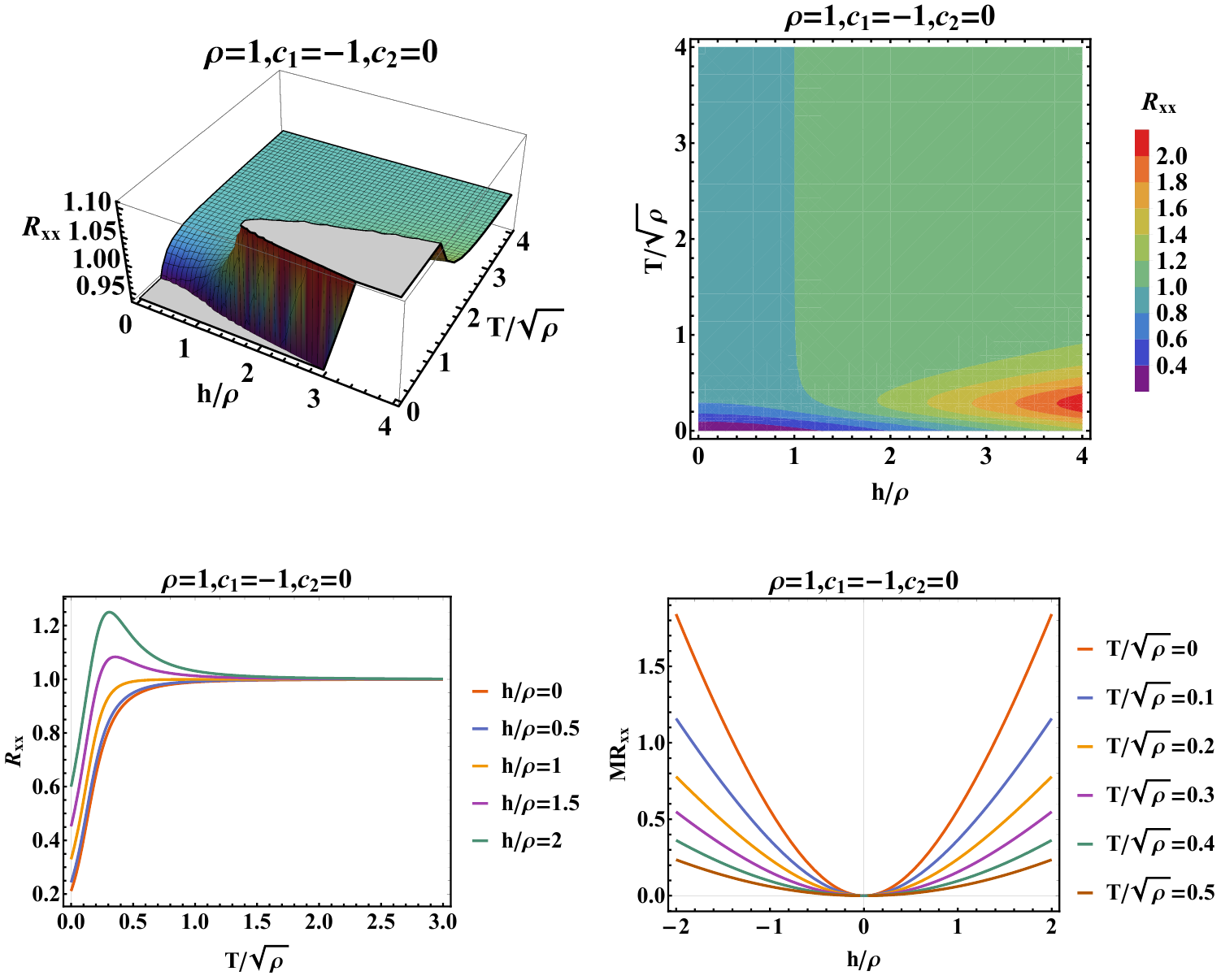}
\caption{Upper left: $R_{xx}$ as a function of $h/\rho$ and $T/\sqrt{\rho}$. Upper right: Contour plot of $R_{xx}$.
Lower left: $R_{xx}$ vs $T/\sqrt{\rho}$ with $h/\rho=0,0.5,1,1.5,2$. Lower right: $MR_{xx}$ vs $h$ with $T/\sqrt{\rho}=0,0.1,0.2,0.3,0.4,0.5$.
We set $T=0$, $c_1=-1$, and $c_2=0$.}
\label{fig:M_Rxx_4}
\end{figure}

Then we present the dependence of $R_{xx}$ on $h/\rho$ and $T/\sqrt{\rho}$ for $c_1=-1$. 
The effect of different $c_1$ and $c_2$ can be deduced by change the temperature correspondingly, 
based on the analysis above. For $h<\rho$, \cref{fig:M_Rxx_4} shows that
the temperature dependence of $R_{xx}$ is monotonic, and corresponds to metallic
behavior. For $h>\rho$, the $R_{xx}$ increases first and then decreases monotonically 
after reaching a maximum. The insulating behavior appears at high temperatures in this case. 
Moreover, if we take larger $c_i$, the metallic behaviors at $h>\rho$ would disappear and the $R_{xx}$
decreases monotonically with increasing temperature. So we come to the conclusion that 
increasing the magnetic field would induce a finite-temperature 
transition or crossover from metallic to insulating behavior.
In the last sub-figure of \cref{fig:M_Rxx_4} the influence of temperature on the 
magneto-resistivity is shown. As temperature increases, the magneto-resistivity
is remarkably suppressed.

\begin{figure}[H]
\centering
\subfigure{\begin{minipage}[t]{0.45\textwidth}
\centering
\includegraphics[scale=0.3]{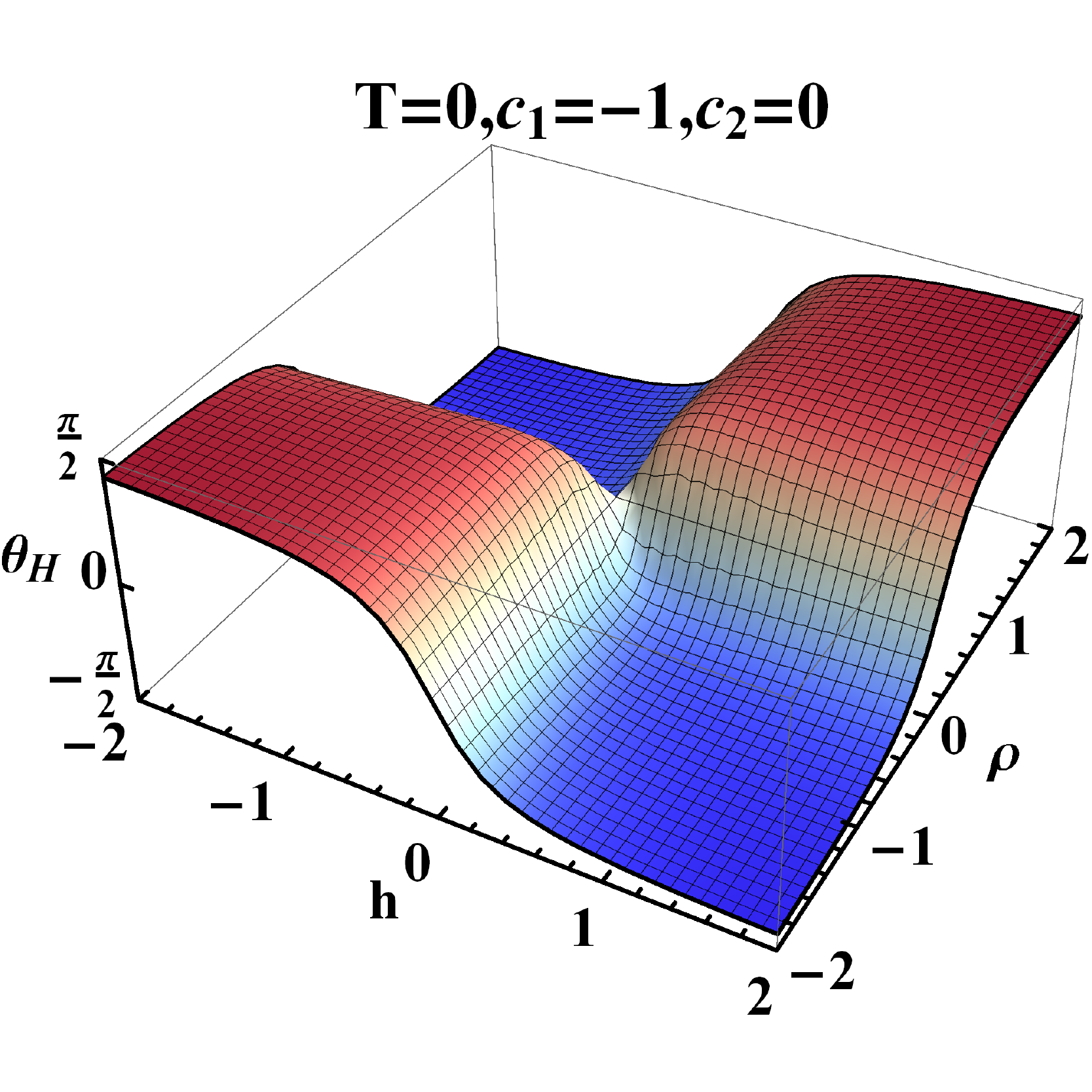}
\end{minipage}
} 
\subfigure{\begin{minipage}[t]{0.45\textwidth}
\centering
\includegraphics[scale=0.3]{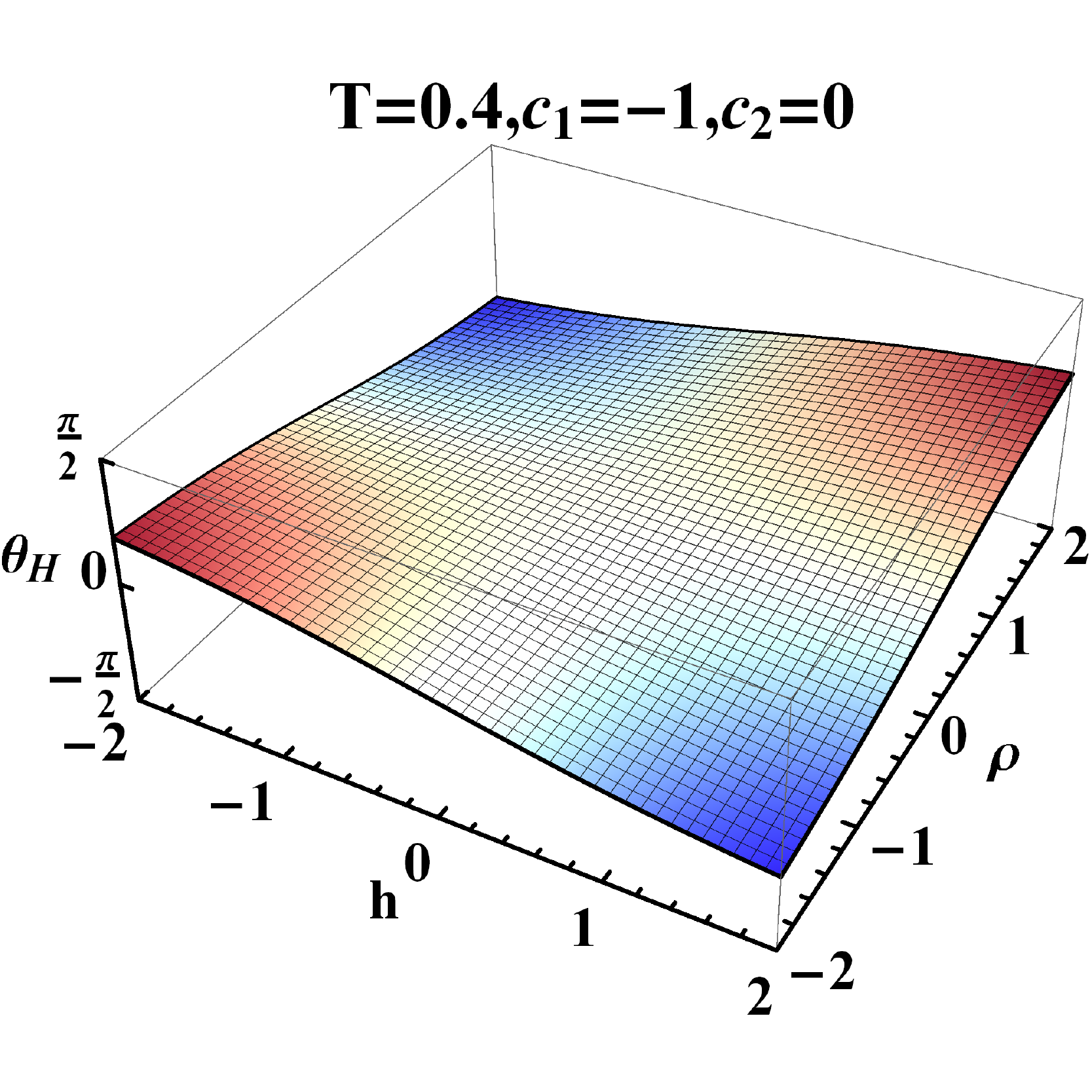}
\end{minipage}
}
\caption{Hall angel at $T=0$ (left panel) and $T=0.4$ (right panel). $c_1=-1$ and $c_2=0$.}
\label{fig:M_theta}
\end{figure}

At the end of the discussion of Maxwell electrodynamics, the Hall angle $\theta_{H}$, defined as 
\begin{equation}
\theta_{H} = \arctan\frac{\sigma_{xy}}{\sigma_{xx}} ,
\end{equation}
is shown in \cref{fig:M_theta}. At zero temperature for large $h$ or $\rho$ the Hall angle get saturated, 
with the sign depend on those of $h$ or $\rho$. The surface is anti-symmetric under the transformation
$\left(\rho,h\right)  \to \left(\rho,-h\right)$ and $\left(\rho,h\right) \to \left(-\rho,h\right)$.
As we expect, the temperature would evidently suppress the Hall angle.

\subsection{Maxwell-Chern-Simons Electrodynamics}
\label{subsec:MCSE}

\begin{figure}[H]
\centering
\includegraphics[width=14cm]{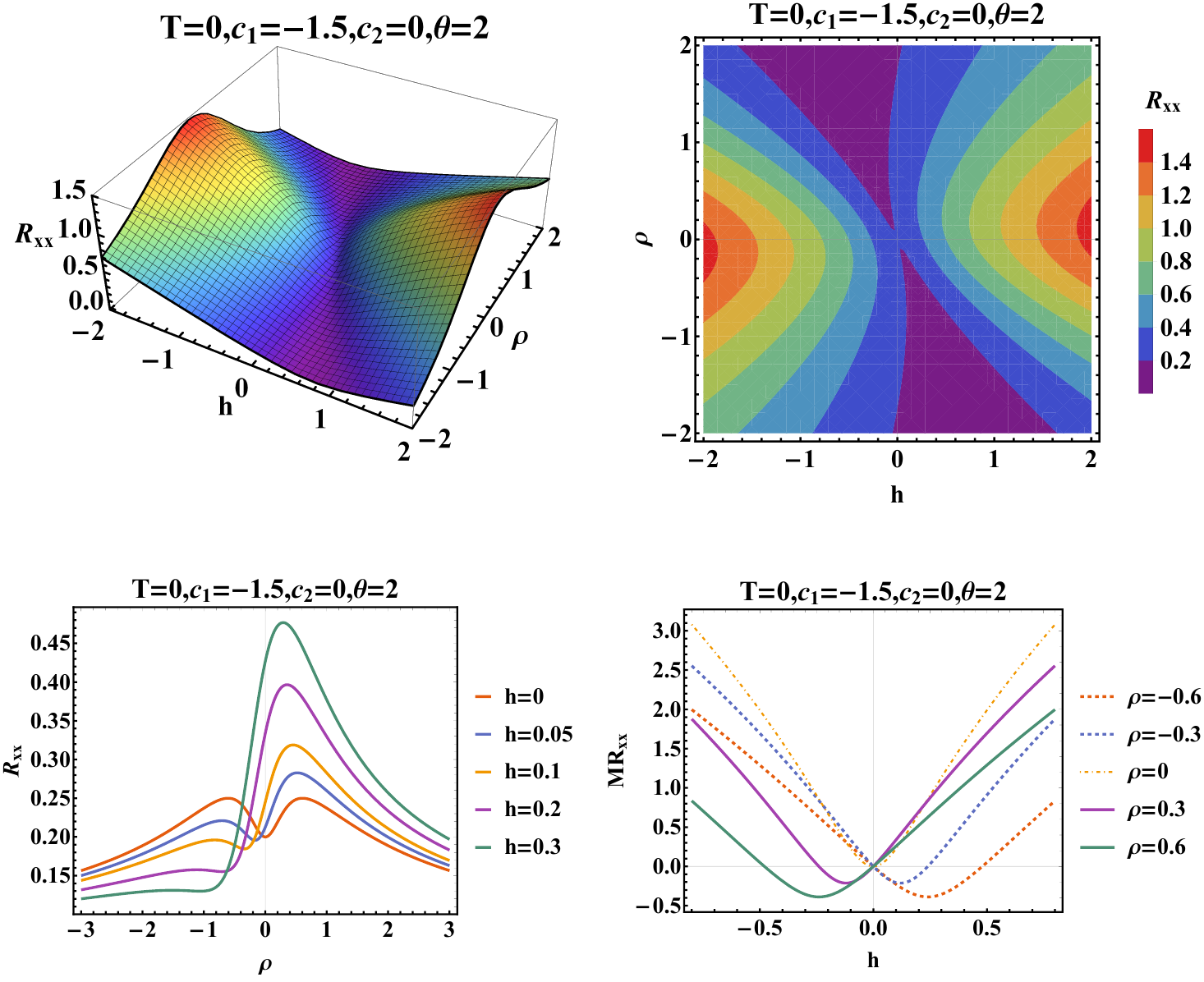}
\caption{Upper left: $R_{xx}$ as a function of $h$ and $\rho$. Upper right: Contour plot of $R_{xx}$.
Lower left: $R_{xx}$ vs $\rho$ with $h=0,0.05,0.1,0.2,0.3$. Lower right: $MR_{xx}$ vs $h$ with $\rho=-0.6,-0.3,0,0.3,0.6$.
We set $T=0$, $c_1=-1.5$, $c_2=0$ and $\theta=2$.}
\label{fig:MCS_Rxx_1}
\end{figure}

\begin{figure}[H]
\centering
\includegraphics[width=14cm]{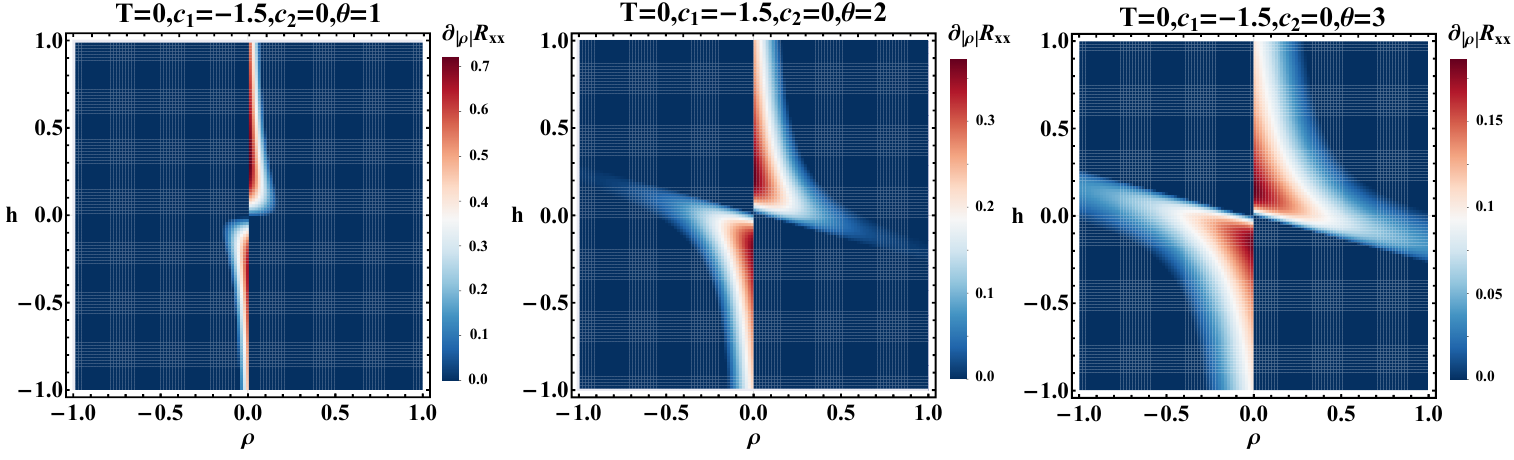}
\caption{Mott-insulating region for $\theta=1,2,3$ in the parameter plane for Maxwell-Chern-Simons 
electrodynamics at zero temperature. We set $c_1=-1.5$ and $c_2=0$.}
\label{fig:MCS_Mott}
\end{figure}

\begin{figure}[H]
\centering
\includegraphics[width=8cm]{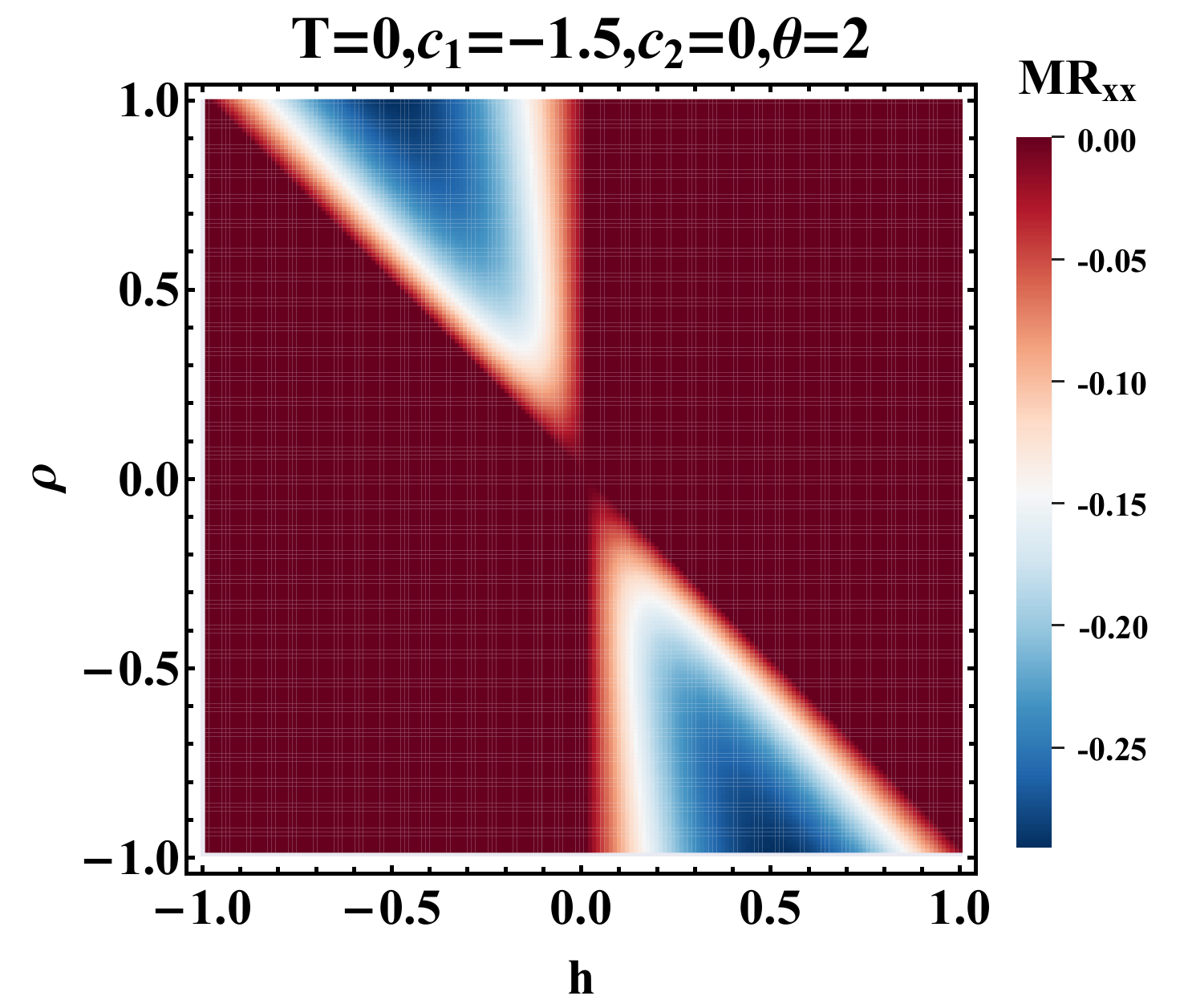}
\caption{Negative magneto-resistivity region for $\theta=3$ in the parameter plane for Maxwell-Chern-Simons 
electrodynamics at zero temperature. We set $c_1=-1.5$ and $c_2=0$.}
\label{fig:MCS_MR}
\end{figure}

\begin{figure}[H]
\centering
\includegraphics[width=14cm]{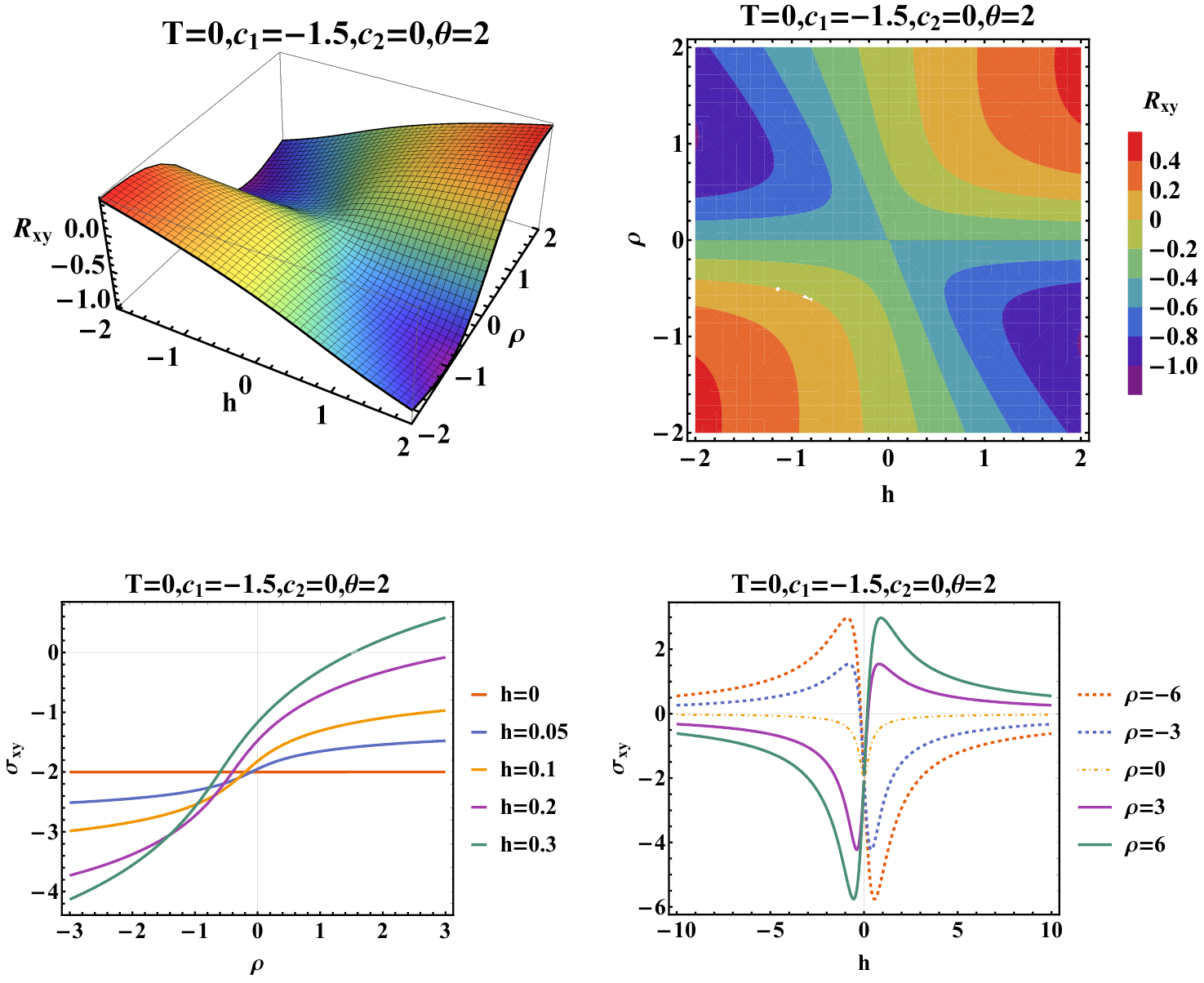}
\caption{Upper left: $R_{xy}$ as a function of $h$ and $\rho$. Upper right: Contour plot of $R_{xy}$.
Lower left: $\sigma_{xy}$ vs $\rho$ with $h=0,0.05,0.1,0.2,0.3$. Lower right: $\sigma_{xy}$ vs $h$ with $\rho=-6,-3,0,3,6$.
We set $T=0$, $c_1=-1.5$, $c_2=0$ and $\theta=2$.}
\label{fig:MCS_Rxy_1}
\end{figure}

The Lorentz and gauge invariance do not forbid the appearance of CP-violating 
Chern-Simons $\theta$ term in the electrodynamics Lagrangian
\begin{equation}
\mathcal{L}(s,p) = s + \theta p,
\end{equation}
and the Chern-Simons theory is of vital importance for the both integer and fractional 
quantum Hall effects in condensed matter physics \cite{Zhang1989:PRL,Fradkin1991:PRB,Zhang1992:IJMPB,DavidTong:QHE}.
And the value of $\theta$ can be related to the Hall conductivity in the unit of $e^2/\hslash$.
After taking the Chern-Simons term into consideration, \cref{eq:rho} gives
\begin{equation}
A_{t}^{\prime}\left(r_{h}\right)=\frac{\rho+\theta h}{r_{h}^{2}} , 
\end{equation} 
and the \cref{eq:temp2} becomes
\begin{equation}
4\pi Tr_{h}-3r_{h}^{2}-c_{1} r_{h}-c_{2} + \frac{\left(1+\theta^{2}\right)h^{2}+2\theta h\rho+\rho^{2}}{4r_{h}^{2}} = 0 .
\end{equation} 
We then reinvestigate the dependence of DC resistivity $R_{xx}$ on $\rho$ and $h$. As one expects, 
the similar saddle surface to that of Maxwell electrodynamics is found, and the reflection asymmetry
and central symmetry due to the $\theta$ term is shown in the upper panel of \cref{fig:MCS_Rxx_1}.
The more surprising things are the appearance of Mott-insulating behavior $\partial_{|\rho|}R_{xx}>0$
and negative magneto-resistivity $MR_{xx}<0$.
At zero field $R_{xx}$ is the even function of $\rho$, and as $|\rho|$ increases, the value of $R_{xx}$
increases and reaches a maximum, showing the feature of Mott-insulating behavior, and then decreases monotonically. 
While for finite magnetic field, the reflection symmetry for positive and negative $\rho$ is broken.
Along the positive $\rho$ direction the behavior of $R_{xx}$ is similar, however, for negative $\rho$
it is more complicated. as $|\rho|$ increases, the value of $R_{xx}$ decreases and reaches a minimum, 
and then share similar behavior as the positive $\rho$ case.
We then further study how the value of $\theta$ affect the Mott-insulating behavior in \cref{fig:MCS_Mott}.
To clearly show the region $\partial_{|\rho|}R_{xx}>0$ we set other region's value to be zero. Similar
results have been obtained as in Ref. \cite{Peng2018:EPJC}. At zero field there is no Mott-insulating
behavior for $\theta=1$. A larger $\theta$ makes it possible even with $h=0$. On the other hand, the
Mott-insulating region becomes larger but $\partial_{|\rho|}R_{xx}$ becomes smaller as $\theta$ increases.
The negative magneto-resistivity revealed in \cref{fig:MCS_Rxx_1} is also shown in the $\rho-h$
parameter plane in \cref{fig:MCS_MR}, from which we can see that the negative magneto-resistivity
emerges in the finite interval of $h/\rho$ of about $\left[-1.2,0\right]$. The positive magneto-resistivity
region's value is set to be zero as well. At larger magnetic field, the negative magneto-resistivity 
phenomenon disappears, and the magneto-resistivity increases almost linearly with $h$. For zero density
the negative magneto-resistivity does not occur.
Finally we show the behavior of Hall resistivity $R_{xy}$ and Hall conductivity $\sigma_{xy}$ in \cref{fig:MCS_Rxy_1}.
The negative Hall resistivity comes from the negative transverse conductivity $\sigma_{xy}$ depending on the direction
of magnetic field $h$ and the type of charge carriers. From the figure we see that at zero field the transverse conductivity
$\sigma_{xy}$ is exactly $-\theta$ in regardless of charge density $\rho$, which agrees with the scenario in quantum Hall 
physics \cite{Zhang1992:IJMPB,Qi2011:RMP}. At finite magnetic field, for positive $\rho$ the Hall conductivity will decrease
to zero, and then changes its sign, while for negative $\rho$ the Hall conductivity remains negative.
And we can also study the impact of magnetic field on the Hall conductivity with fixed $\rho$. All the transverse conductivity
for various $\rho$ as strong field tend to be zero as the result of localization. For positive $\rho$, in the positive
magnetic field, the Hall conductivity decreases to zero, then changes its sign and increases to a maximum, and finally 
decreases monotonically to zero from the positive side. And in the negative magnetic field, the Hall conductivity increases 
to a negative maximum, and then decreases to zero from the negative side. The case of negative $\rho$ can be known by simply
making transformation $\left(\rho,h\right)\to\left(-\rho,-h\right)$ in the discussion above.

\subsection{Born-Infeld Electrodynamics}

The Born-Infeld electrodynamics is described by a square-root Lagrangian \cite{Born1933:Nat,Born1934:PRSLA}
\begin{equation}
\mathcal{L}(s,p)=\frac{1}{a}\left(1-\sqrt{1-2as-a^{2}p^{2}}\right) ,
\end{equation}
where the coupling parameter $a=\left(2\pi\alpha'\right)^{2}$ relates to the Regge slope $\alpha'$. 
It is believed that such a NLED governs the dynamics of electromagnetic fields on D-branes. 
If we take the zero-slope limit $\alpha'\to0$, the Maxwell Lagrangian is recovered
\begin{equation}
\mathcal{L}(s,p)=s+\mathcal{O}(a) .
\label{eq:BI_ex}
\end{equation}
The Born-Infeld electrodynamics takes the advantage of eliminating the divergence of electrostatic self-energy
and incorporating maximal electric fields \cite{zwiebach2004:CUP}. This can be seen from the solution of
\cref{eq:rho}
\begin{equation}
A_{t}^{\prime}\left(r\right)=\frac{\rho}{\sqrt{a\left(\rho^{2}+h^{2}\right)+r^{4}}} ,
\label{eq:E_BI}
\end{equation}
which is finite when $r\to 0$. The $r_h$ is solved form 
\begin{equation}
4\pi r_{h}T-3r_{h}^{2}-c_{1}r_{h}-c_{2}+\frac{\rho^{2}}{2\sqrt{a\left(h^{2}+\rho^{2}\right)+r_{h}^{4}}}+
\frac{1}{2a}\left(\sqrt{\frac{\left(ah^{2}+r_{h}^{4}\right)^{2}}{a\left(h^{2}+\rho^{2}\right)+r_{h}^{4}}}-r_{h}^{2}\right)=0 .
\end{equation}
\begin{figure}[H]
\centering
\includegraphics[width=14cm]{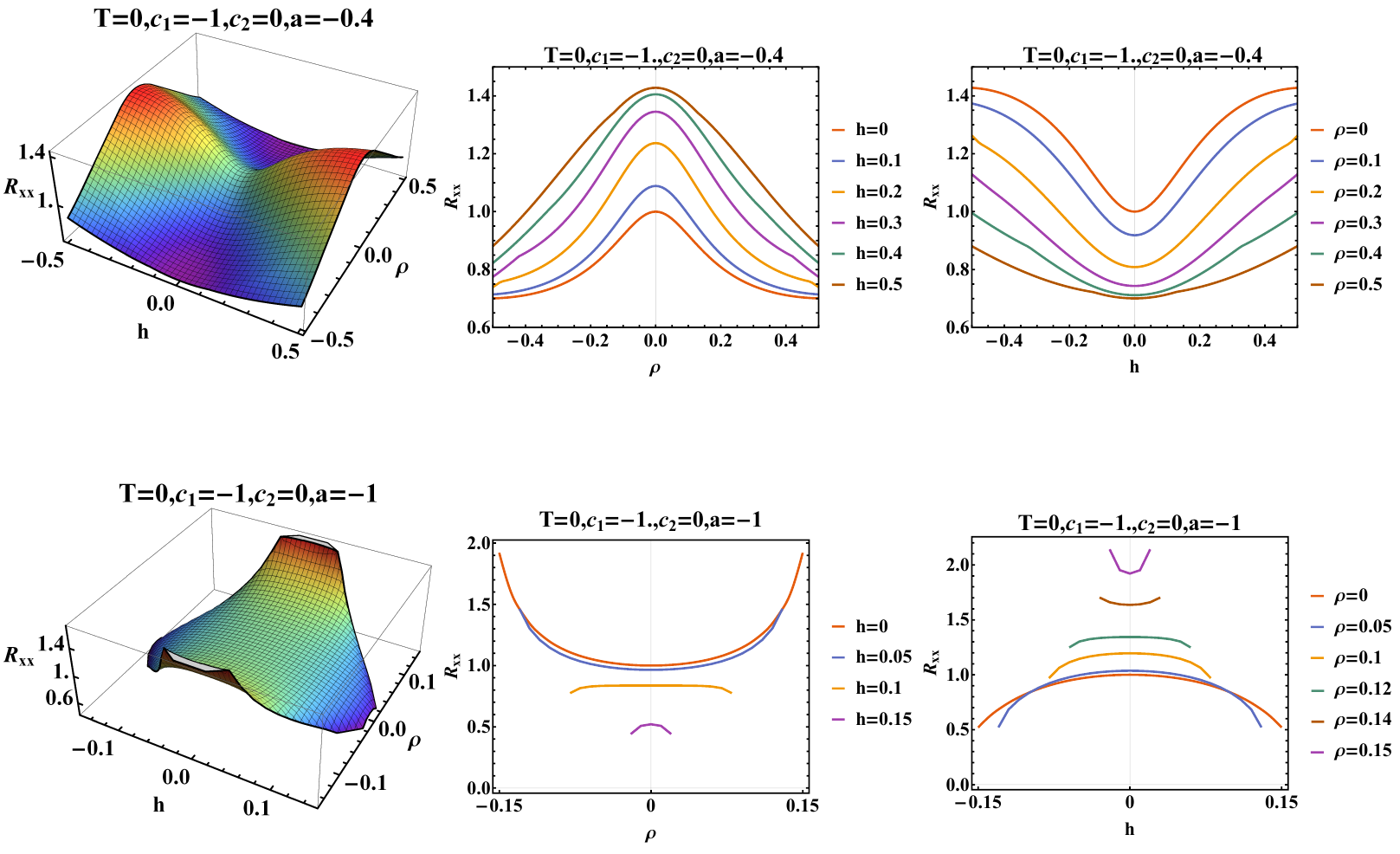}
\caption{$R_{xx}$ as a function of $h$ and $\rho$ and some intersection curves 
for fixed $h$ and $\rho$ for $a=-0.4$ (upper panel) and $a=-1$ (lower panel). 
The surface ends at some place where $h$ and $\rho$ touch the upper bound.}
\label{fig:BI_Rxx}
\end{figure}

\begin{figure}[H]
\centering
\includegraphics[width=14cm]{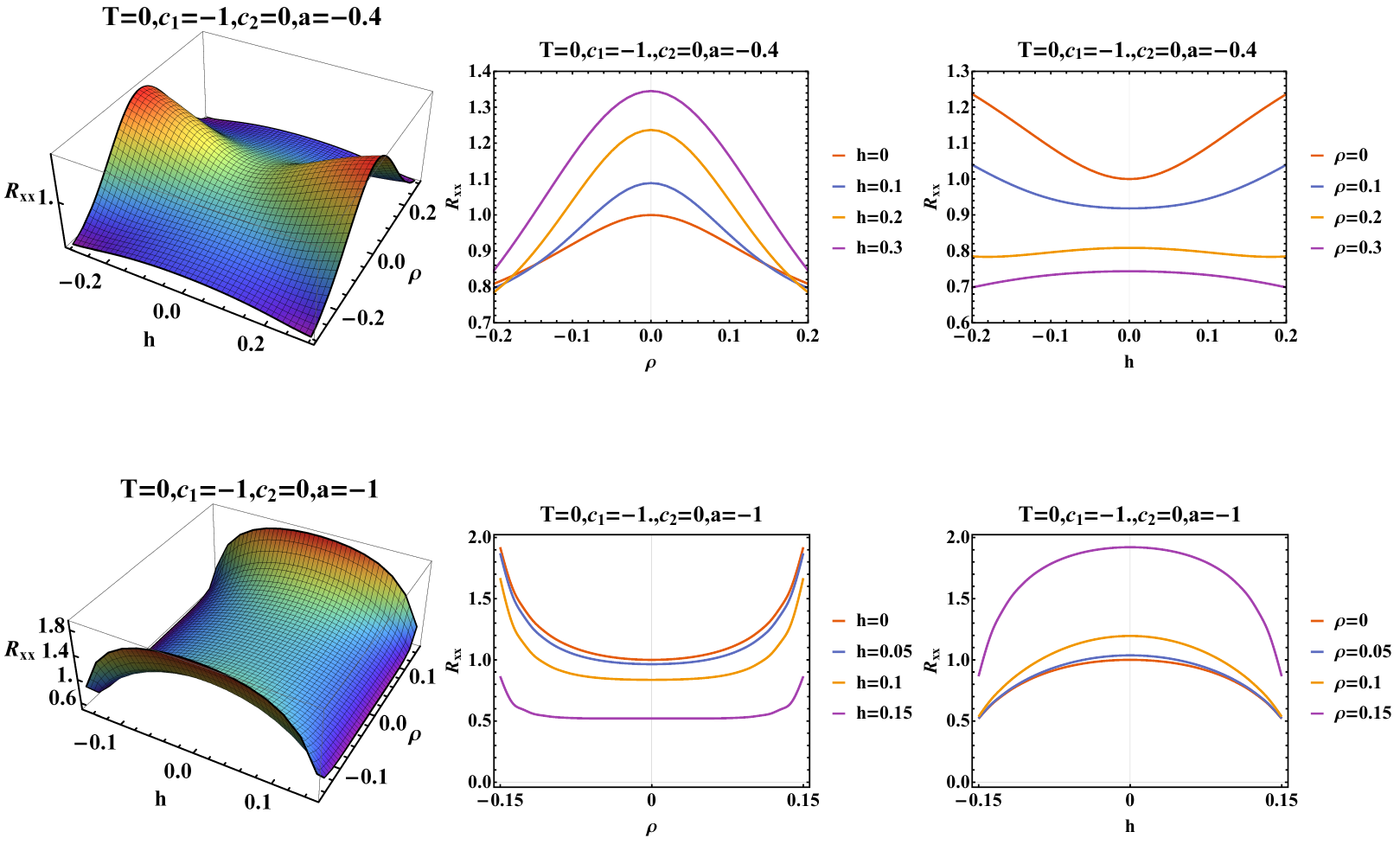}
\caption{$R_{xy}$ as a function of $h$ and $\rho$ and some intersection curves 
for fixed $h$ and $\rho$ for $a=-0.4$ (upper panel) and $a=-1$ (lower panel).}
\label{fig:S_Rxx}
\end{figure}

The DC resistivity is then studied in the framework of Born-Infeld electrodynamics 
and the results are shown in \cref{fig:S_Rxx}. For positive $a$ the behavior of $R_{xx}$ 
is similar to the Maxwell case \cite{Peng2018:EPJC,Cremonini2017:JHEP}, and the negative 
$a$ can bring in more interesting phenomenon. However, for negative $a$ the \cref{eq:E_BI} 
suffers a singularity $r=r_{s}\equiv\sqrt[1/4]{\left|a\right|\left(h^{2}+\rho^{2}\right)}$,
and the physical solution requires that $r_{h}>r_{s}$, setting an upper bound for 
$h^{2}+\rho^{2}$ \cite{Peng2018:EPJC}. We then present two cases of $a=-0.4$ and $a=-1$, 
respectively. For $a=-0.4$, We the saddle surface is similar to the previous results,
and there is no Mott-insulating behaviors and negative magneto-resistivity. However,
if one increases the absolute value of the negative $a$, the region admitting physical 
solution for $R_{xx}$ shrinks and is truncated at the upper bound of $h^{2}+\rho^{2}$.
For zero and finite small magnetic field, the Mott-insulating behavior is absent, while
for a larger field the Mott-insulating behavior emerges. And for a relatively small
$\rho$, the negative magneto-resistivity is present, and increasing $\rho$ can destroy
it. To expose the role of $a$ we expand the interaction between electrons to the first
order \cite{Peng2018:EPJC}
\begin{equation}
F(a)=\rho A_{t}^{\prime}\left(r\right)\sim\frac{\rho^{2}}{r^{2}}\left[1
-\frac{a}{2r^{4}}\left(h^{2}+\rho^{2}\right)\right]+\mathcal{O}(a^{2}) ,
\end{equation}
in which the leading order is the familiar Coulomb interaction, and the nonlinearity parameter
serves as an effective modification. A positive $a$ suppress the interaction and we do not 
expect different phenomenon compared with the Maxwell electrodynamics. However, for a negative
$a$ the interaction is enhanced at $r_h$
\begin{equation}
\frac{F(a)}{F(0)}=1+\frac{1}{2}\left(\frac{r_{s}}{r_{h}}\right)^{4} ,
\end{equation}
and we expect that the NLED model with negative $a$ can grasp some features of strongly correlated systems.
Besides, in the leading order expansion \cref{eq:BI_ex} the Chern-Simons term does not appear,
and can be deduced from \cref{eq:Rxx_2} that the system will not exhibit Mott-insulating behavior
and negative magneto-resistivity. Thus we conclude that the Born-Infeld electrodynamics provides a
new mechanism different from the Chern-Simons theory to give rise to the Mott-insulating behavior
and negative magneto-resistivity, and the temperature can induce a transition at finite temperature.

One can also consider another kind of construction of square-root Lagrangian \cite{zwiebach2004:CUP}
\begin{equation}
\mathcal{L}(s,p)=\frac{1}{a}\left(1-\sqrt{1-2as}\right) ,
\end{equation}
which has the same leading order expansion as \cref{eq:BI_ex}, and will re-derive the results of 
Born-Infeld electrodynamics in the presence of zero field. The solution of \cref{eq:rho} gives
\begin{equation}
A_{t}^{\prime}\left(r\right)=\frac{\rho}{r^{2}}\sqrt{\frac{ah^{2}+r^{4}}{a\rho^{2}+r^{4}}} ,
\end{equation}
which can be deduced that the effective interaction is 
\begin{equation}
F=\rho A_{t}^{\prime}\left(r\right)\sim\frac{\rho^{2}}{r^{2}}\left[1
-\frac{a}{2r^{4}}\left(\rho^{2}-h^{2}\right)\right]+\mathcal{O}(a^{2}) .
\end{equation}
The radius equation is
\begin{equation}
4\pi r_{h}T-3r_{h}^{2}-c_{1}r_{h}-c_{2}+\frac{h^{2}}{2r_{h}^{2}}+\frac{r_{h}^{2}}{2a}
\left(\sqrt{\frac{a\rho^{2}+r_{h}^{4}}{ah^{2}+r_{h}^{4}}}-1\right)=0 .
\end{equation}

We re-check the $R_{xx}$ in the square-root electrodynamics for negative $a$. 
For $a=-0.4$ the saddle surface is similar to the previous results, and both 
the zero field and finite field no Mott insulating behaviors are shown. 
And for fixed $\rho$, it is surprising that for larger $\rho$,
the transition from positive magneto-resistivity to negative magneto-resistivity
is observed for about $\rho>h$. We then studied the case of larger	negative $a$. 
To our surprise, the saddle surface changes its direction and looks as if it rotates $90^\circ$.
We find that the Mott-insulating behavior appears and becomes significant at large
$\rho$, and negative magneto-resistivity is observed for various $\rho$.

\section{Conclusion}
\label{sec:Conc}

In this work the black brane solution of four-dimensional massive gravity with 
backreacted NLED is obtained, and with the dictionary of gauge/gravity duality,
the transport properties of the strongly correlated systems in the presence of 
finite magnetic field in 2+1-dimensional boundary is studied. In our holographic 
setup the bulk geometry and NLED field are perturbed, and the DC conductivities 
are obtained in the linear response regime. Then some general properties are 
obtained in various limit, which agrees well with the previous work. To make 
it concrete, we present the study of the conventional Maxwell electrodynamics,
the topological non-trivial Maxwell-Chern-Simons electrodynamics, and the 
Born-Infeld electrodynamics with string-theoretical correction taken into
consideration. We concentrate on two interesting phenomena, i.e., the Mott-insulating 
behavior and negative magneto-resistivity, and results at zero temperature are 
summarized in \cref{tab:conclusion}.

\begin{table}[htbp]
  \centering
    \begin{tabular}{|c|c|c|c|c|} 
    \hline
          & Lagrangian & Parameter & \multicolumn{1}{c|}{Mott-insulating behavior} & Negative magneto-resistivity \\
    \hline
    Maxwell & $s$   &       & \multicolumn{1}{c|}{No} & No \\
    \hline
    Maxwell-& \multirow{3}[2]{*}{$s+\theta p$} & \multirow{3}[2]{*}{$\theta$} & \multirow{3}[2]{*}{See \cref{fig:MCS_Mott}} & \multirow{3}[2]{*}{See \cref{fig:MCS_MR}} \\
    Chern- &       &       &       &  \\
    Simons &       &       &       &  \\
    \hline
    \multirow{4}[6]{*}{Born-Infeld} & \multirow{4}[6]{*}{$\dfrac{1}{a}\left(1-\sqrt{1-2as-a^2 p^2}\right)$} & $a>0$ & \multicolumn{1}{c|}{No} & \multicolumn{1}{c|}{No} \\
	\cline{3-5}          &       & $a=-0.4$ & \multicolumn{1}{c|}{No} & \multicolumn{1}{c|}{No} \\
	\cline{3-5}          &       & \multirow{2}[2]{*}{$a=-1$} & \multicolumn{1}{c|}{Finite $h$ and small $\rho$.} & \multicolumn{1}{c|}{Finite $h$ and small $\rho$.} \\
          &       &       &  See \cref{fig:BI_Rxx} & See \cref{fig:BI_Rxx} \\
    \hline
    \multirow{3}[6]{*}{Square} & \multirow{3}[6]{*}{$\dfrac{1}{a}\left(1-\sqrt{1-2as}\right)$} & $a>0$ & \multicolumn{1}{c|}{No} & No \\
	\cline{3-5}          &       & $a=-0.4$ & \multicolumn{1}{c|}{No} & \multicolumn{1}{c|}{Larger $\rho$. See \cref{fig:S_Rxx}} \\
	\cline{3-5}          &       & $a=-1$ & Larger $\rho$. See \cref{fig:S_Rxx} & Yes \\
    \hline
    \end{tabular}
  \caption{Conclusion of Mott-insulating behavior and negative magneto-resistivity for various NLED models at zero temperature $T=0$.}
  \label{tab:conclusion}
\end{table}

The massive gravity coupling parameters' influence on the in-plane resistivity 
is compared with the temperature, where we find the $c_{1}$ behaves as the effective
temperature correction, and non-zero one can lead to the DC conductivity. While 
the $c_{2}$ has less significant effect on $R_{xx}$ than $c_{1}$. Moreover, the 
dependence on $\rho$ and $h$ is shown and we find the field can induce the metal-insulator
transition or Mott-insulating behavior and negative magneto-resistivity.
Two different mechanism, the Chern-Simons term, and the negative nonlinearity parameter
are proved that can give rise to Mott-insulating behavior and negative magneto-resistivity.
We hope our work could explain some experimental phenomenon in strongly correlated systems.

\section*{Acknowledgment}

We are grateful to thank Peng Wang for useful discussions. 
This work is supported by NSFC (Grant No.11947408).

\bibliographystyle{JHEP}
\bibliography{ref}

\providecommand{\href}[2]{#2}\begingroup\raggedright\begin{thebibliography}{10}

\bibitem{Susskind1997:PRD}
T.~Banks, W.~Fischler, S.H.~Shenker and L.~Susskind, \emph{{$M$} theory as a
  matrix model: A conjecture},
  \href{https://doi.org/10.1103/PhysRevD.55.5112}{\emph{Phys. Rev. D}
  {\bfseries 55} (1997) 5112}
  [\href{https://arxiv.org/abs/hep-th/9610043}{{\ttfamily hep-th/9610043}}].

\bibitem{Maldacena1997:IJTP}
J.~Maldacena, \emph{The large-{$N$} limit of superconformal field theories and
  supergravity}, \href{https://doi.org/10.1023/A:1026654312961}{\emph{Int. J.
  Theor. Phys.} {\bfseries 38} (1999) 1113}
  [\href{https://arxiv.org/abs/hep-th/9711200}{{\ttfamily hep-th/9711200}}].

\bibitem{Witten1998:ATMP}
E.~Witten, \emph{Anti-de sitter space and holography},
  \href{https://doi.org/10.4310/ATMP.1998.v2.n2.a2}{\emph{Adv. Theor. Math.
  Phys.} {\bfseries 2} (1998) }
  [\href{https://arxiv.org/abs/hep-th/9802150}{{\ttfamily hep-th/9802150}}].

\bibitem{Ammon2015:CUP}
M.~Ammon and J.~Erdmenger, \emph{Gauge/Gravity Duality: Foundations and
  Applications}, Cambridge University Press (2015).

\bibitem{Eleftherios2011:LNP}
E.~Papantonopoulos, \emph{From Gravity to Thermal Gauge Theories: The
  {A}d{S}/{CFT} Correspondence}, vol.~828 of \emph{Lecture Notes in Physics},
  Springer-Verlag Berlin Heidelberg (2011),
  \href{https://doi.org/10.1007/978-3-642-04864-7}{10.1007/978-3-642-04864-7}.

\bibitem{Natsuume2016:LNP}
M.~Natsuume, \emph{AdS/CFT Duality User Guide}, vol.~903 of \emph{Lecture Notes
  in Physics}, Springer (2015),
  \href{https://doi.org/10.1007/978-4-431-55441-7}{10.1007/978-4-431-55441-7},
  [\href{https://arxiv.org/abs/1409.3575}{{\ttfamily 1409.3575}}].

\bibitem{Policastro2001:PRL}
G.~Policastro, D.T.~Son and A.O.~Starinets, \emph{Shear viscosity of strongly
  coupled $\mathcal{N}=4$ supersymmetric {Y}ang-{M}ills plasma},
  \href{https://doi.org/10.1103/PhysRevLett.87.081601}{\emph{Phys. Rev. Lett.}
  {\bfseries 87} (2001) 081601}
  [\href{https://arxiv.org/abs/hep-th/0104066}{{\ttfamily hep-th/0104066}}].

\bibitem{Buchel2004:PRL}
A.~Buchel and J.T.~Liu, \emph{Universality of the shear viscosity from
  supergravity duals},
  \href{https://doi.org/10.1103/PhysRevLett.93.090602}{\emph{Phys. Rev. Lett.}
  {\bfseries 93} (2004) 090602}
  [\href{https://arxiv.org/abs/hep-th/0311175}{{\ttfamily hep-th/0311175}}].

\bibitem{Kovtun2005:PRL}
P.K.~Kovtun, D.T.~Son and A.O.~Starinets, \emph{Viscosity in strongly
  interacting quantum field theories from black hole physics},
  \href{https://doi.org/10.1103/PhysRevLett.94.111601}{\emph{Phys. Rev. Lett.}
  {\bfseries 94} (2005) 111601}
  [\href{https://arxiv.org/abs/hep-th/0405231}{{\ttfamily hep-th/0405231}}].

\bibitem{Benincasa2006:JHEP}
P.~Benincasa and A.~Buchel, \emph{Transport properties of $\mathcal{N}=4$
  supersymmetric {Y}ang-{M}ills theory at finite coupling},
  \href{https://doi.org/10.1088/1126-6708/2006/01/103}{\emph{Journal of High
  Energy Physics} {\bfseries 2006} (2006) 103}
  [\href{https://arxiv.org/abs/hep-th/0510041}{{\ttfamily hep-th/0510041}}].

\bibitem{DTSon2006:JHEP}
D.T.~Son and A.O.~Starinets, \emph{Hydrodynamics of {$R$}-charged black holes},
  \href{https://doi.org/10.1088/1126-6708/2006/03/052}{\emph{Journal of High
  Energy Physics} {\bfseries 2006} (2006) 052}
  [\href{https://arxiv.org/abs/hep-th/0601157}{{\ttfamily hep-th/0601157}}].

\bibitem{DTSon2007:ARNPS}
D.T.~Son and A.O.~Starinets, \emph{Viscosity, black holes, and quantum field
  theory},
  \href{https://doi.org/10.1146/annurev.nucl.57.090506.123120}{\emph{Annu. Rev.
  Nucl. Part. Sci.} {\bfseries 57} (2007) 95}
  [\href{https://arxiv.org/abs/0704.0240}{{\ttfamily 0704.0240}}].

\bibitem{Kovtun2007:JPA}
P.~Kovtun, \emph{Lectures on hydrodynamic fluctuations in relativistic
  theories}, \href{https://doi.org/10.1088/1751-8113/45/47/473001}{\emph{J.
  Phys. A: Math. Theor.} {\bfseries 45} (2012) 473001}
  [\href{https://arxiv.org/abs/1205.5040}{{\ttfamily 1205.5040}}].

\bibitem{Rangamani2009:CQG}
M.~Rangamani, \emph{Gravity and hydrodynamics: Lectures on the fluid-gravity
  correspondence},
  \href{https://doi.org/10.1088/0264-9381/26/22/224003}{\emph{Class. Quant.
  Grav.} {\bfseries 26} (2009) 224003}
  [\href{https://arxiv.org/abs/0905.4352}{{\ttfamily 0905.4352}}].

\bibitem{Adams2001:PRD}
A.~Adams and E.~Silverstein, \emph{Closed string tachyons, {A}d{S}/{CFT}, and
  large {$N$} {QCD}},
  \href{https://doi.org/10.1103/PhysRevD.64.086001}{\emph{Phys. Rev. D}
  {\bfseries 64} (2001) 086001}
  [\href{https://arxiv.org/abs/hep-th/0103220}{{\ttfamily hep-th/0103220}}].

\bibitem{Brodsky2004:PLB}
S.J.~Brodsky and G.F.~De~T{\'e}ramond, \emph{Light-front hadron dynamics and
  {A}d{S}/{CFT} correspondence},
  \href{https://doi.org/10.1016/j.physletb.2003.12.050}{\emph{Phys. Lett. B}
  {\bfseries 582} (2004) 211}
  [\href{https://arxiv.org/abs/hep-th/0310227}{{\ttfamily hep-th/0310227}}].

\bibitem{Brodsky2005:PRL}
G.F.~de~T\'eramond and S.J.~Brodsky, \emph{Hadronic spectrum of a holographic
  dual of {QCD}},
  \href{https://doi.org/10.1103/PhysRevLett.94.201601}{\emph{Phys. Rev. Lett.}
  {\bfseries 94} (2005) 201601}
  [\href{https://arxiv.org/abs/hep-th/0501022}{{\ttfamily hep-th/0501022}}].

\bibitem{Brodsky2009:PRL}
G.F.~de~T\'eramond and S.J.~Brodsky, \emph{Light-front holography: A first
  approximation to {QCD}},
  \href{https://doi.org/10.1103/PhysRevLett.102.081601}{\emph{Phys. Rev. Lett.}
  {\bfseries 102} (2009) 081601}
  [\href{https://arxiv.org/abs/0809.4899}{{\ttfamily 0809.4899}}].

\bibitem{Erlich2005:PRL}
J.~Erlich, E.~Katz, D.T.~Son and M.A.~Stephanov, \emph{{QCD} and a holographic
  model of hadrons},
  \href{https://doi.org/10.1103/PhysRevLett.95.261602}{\emph{Phys. Rev. Lett.}
  {\bfseries 95} (2005) 261602}
  [\href{https://arxiv.org/abs/hep-ph/0501128}{{\ttfamily hep-ph/0501128}}].

\bibitem{Rold20005:NPB}
L.~Da~Rold and A.~Pomarol, \emph{Chiral symmetry breaking from five-dimensional
  spaces}, \href{https://doi.org/10.1016/j.nuclphysb.2005.05.009}{\emph{Nucl.
  Phys. B} {\bfseries 721} (2005) 79}
  [\href{https://arxiv.org/abs/hep-ph/0501218}{{\ttfamily hep-ph/0501218}}].

\bibitem{Zayakin2008:JHEP}
A.~Zayakin, \emph{{QCD} vacuum properties in a magnetic field from
  {A}d{S}/{CFT}: Chiral condensate and goldstone mass},
  \href{https://doi.org/10.1088/1126-6708/2008/07/116}{\emph{Journal of High
  Energy Physics} {\bfseries 2008} (2008) 116}
  [\href{https://arxiv.org/abs/0807.2917}{{\ttfamily 0807.2917}}].

\bibitem{Edelstein2009:AIP}
J.D.~Edelstein, J.P.~Shock and D.~Zoakos, \emph{The {A}d{S}/{CFT}
  correspondence and non-perturbative {QCD}},
  \href{https://doi.org/10.1063/1.3131566}{\emph{AIP Conf. Proc.} {\bfseries
  1116} (2009) 265} [\href{https://arxiv.org/abs/0901.2534}{{\ttfamily
  0901.2534}}].

\bibitem{Gursoy2011:QCD}
U.~Gursoy, E.~Kiritsis, L.~Mazzanti, G.~Michalogiorgakis and F.~Nitti,
  \emph{Improved holographic {QCD}},  in \emph{From Gravity to Thermal Gauge
  Theories: The AdS/CFT Correspondence}, pp.~79--146, Springer (2011),
  \href{https://doi.org/10.1007/978-3-642-04864-7_4}{DOI}
  [\href{https://arxiv.org/abs/1006.5461}{{\ttfamily 1006.5461}}].

\bibitem{Alfimov2015:JHEP}
M.~Alfimov, N.~Gromov and V.~Kazakov, \emph{{QCD} pomeron from {A}d{S}/{CFT}
  quantum spectral curve},
  \href{https://doi.org/10.1007/JHEP07(2015)164}{\emph{Journal of High Energy
  Physics} {\bfseries 2015} (2015) 164}
  [\href{https://arxiv.org/abs/1408.2530}{{\ttfamily 1408.2530}}].

\bibitem{Bergman2007:JHEP}
O.~Bergman, M.~Lippert and G.~Lifschytz, \emph{Holographic nuclear physics},
  \href{https://doi.org/10.1088/1126-6708/2007/11/056}{\emph{Journal of High
  Energy Physics} {\bfseries 2007} (2007) 056}
  [\href{https://arxiv.org/abs/0708.0326}{{\ttfamily 0708.0326}}].

\bibitem{Baldino2017:PRD}
S.~Baldino, S.~Bolognesi, S.B.~Gudnason and D.~Koksal, \emph{Solitonic approach
  to holographic nuclear physics},
  \href{https://doi.org/10.1103/PhysRevD.96.034008}{\emph{Phys. Rev. D}
  {\bfseries 96} (2017) 034008}
  [\href{https://arxiv.org/abs/1703.08695}{{\ttfamily 1703.08695}}].

\bibitem{Hartnoll2008:PRL}
S.A.~Hartnoll, C.P.~Herzog and G.T.~Horowitz, \emph{Building a holographic
  superconductor},
  \href{https://doi.org/10.1103/PhysRevLett.101.031601}{\emph{Phys. Rev. Lett.}
  {\bfseries 101} (2008) 031601}
  [\href{https://arxiv.org/abs/0803.3295}{{\ttfamily 0803.3295}}].

\bibitem{Hartnoll2009:CQG}
S.A.~Hartnoll, \emph{Lectures on holographic methods for condensed matter
  physics}, \href{https://doi.org/10.1088/0264-9381/26/22/224002}{\emph{Class.
  Quant. Grav.} {\bfseries 26} (2009) 224002}
  [\href{https://arxiv.org/abs/0903.3246}{{\ttfamily 0903.3246}}].

\bibitem{Herzog2009:JPA}
C.P.~Herzog, \emph{Lectures on holographic superfluidity and
  superconductivity},
  \href{https://doi.org/10.1088/1751-8113/42/34/343001}{\emph{Journal of
  Physics A: Mathematical and Theoretical} {\bfseries 42} (2009) 343001}
  [\href{https://arxiv.org/abs/0904.1975}{{\ttfamily 0904.1975}}].

\bibitem{Herzog2010:PRD}
C.P.~Herzog, \emph{Analytic holographic superconductor},
  \href{https://doi.org/10.1103/PhysRevD.81.126009}{\emph{Phys. Rev. D}
  {\bfseries 81} (2010) 126009}
  [\href{https://arxiv.org/abs/1003.3278}{{\ttfamily 1003.3278}}].

\bibitem{Mcgreevy2010:AHEP}
J.~McGreevy, \emph{Holographic duality with a view toward many-body physics},
  \href{https://doi.org/10.1155/2010/723105}{\emph{Adv. High Energy Phys.}
  {\bfseries 2010} (2010) } [\href{https://arxiv.org/abs/0909.0518}{{\ttfamily
  0909.0518}}].

\bibitem{Nishioka2010:JHEP}
T.~Nishioka, S.~Ryu and T.~Takayanagi, \emph{Holographic
  superconductor/insulator transition at zero temperature},
  \href{https://doi.org/10.1007/JHEP03(2010)131}{\emph{Journal of High Energy
  Physics} {\bfseries 2010} (2010) 131}
  [\href{https://arxiv.org/abs/0911.0962}{{\ttfamily 0911.0962}}].

\bibitem{Cubrovic2009:Sci}
M.~{\v{C}}ubrovi{\'c}, J.~Zaanen and K.~Schalm, \emph{String theory, quantum
  phase transitions, and the emergent {F}ermi liquid},
  \href{https://doi.org/10.1126/science.1174962}{\emph{Science} {\bfseries 325}
  (2009) 439} [\href{https://arxiv.org/abs/0904.1993}{{\ttfamily 0904.1993}}].

\bibitem{Liu2011:PRD}
H.~Liu, J.~McGreevy and D.~Vegh, \emph{Non-{F}ermi liquids from holography},
  \href{https://doi.org/10.1103/PhysRevD.83.065029}{\emph{Phys. Rev. D}
  {\bfseries 83} (2011) 065029}
  [\href{https://arxiv.org/abs/0903.2477}{{\ttfamily 0903.2477}}].

\bibitem{Iqbal2012:NFL}
N.~Iqbal, H.~Liu and M.~Mezei, \emph{Lectures on holographic non-{F}ermi
  liquids and quantum phase transitions},  in \emph{String Theory And Its
  Applications: TASI 2010 From meV to the Planck Scale}, pp.~707--815, World
  Scientific (2012), \href{https://doi.org/10.1142/9789814350525_0013}{DOI}
  [\href{https://arxiv.org/abs/1110.3814}{{\ttfamily 1110.3814}}].

\bibitem{Faulkner2011:JHEP}
T.~Faulkner and J.~Polchinski, \emph{Semi-holographic {F}ermi liquids},
  \href{https://doi.org/10.1007/JHEP06(2011)012}{\emph{Journal of High Energy
  Physics} {\bfseries 2011} (2011) 12}
  [\href{https://arxiv.org/abs/1001.5049}{{\ttfamily 1001.5049}}].

\bibitem{Cai2015:SCP}
R.~Cai, L.~Li, L.~Li and R.~Yang, \emph{Introduction to holographic
  superconductor models},
  \href{https://doi.org/10.1007/s11433-015-5676-5}{\emph{Sci. China Phys. Mech.
  Astron.} {\bfseries 58} (2015) 1}
  [\href{https://arxiv.org/abs/1502.00437}{{\ttfamily 1502.00437}}].

\bibitem{AGD2012:QFT}
A.A.~Abrikosov, L.P.~Gorkov and I.E.~Dzyaloshinski, \emph{Methods of Quantum
  Field Theory in Statistical Physics}, Courier Corporation (2012).

\bibitem{Landau:vol9}
E.~Lifshitz and L.P.~Pitaevskii, \emph{Statistical Physics: Theory of the
  Condensed State}, vol.~9 of \emph{Course of Theoretical Physics},
  Butterworth-Heinemann (2013).

\bibitem{Hall1879}
E.H.~Hall, \emph{On a new action of the magnet on electric currents},
  \href{https://doi.org/10.2307/2369245}{\emph{American Journal of Mathematics}
  {\bfseries 2} (1879) 287}.

\bibitem{Klitzing1980:PRL}
K.v.~Klitzing, G.~Dorda and M.~Pepper, \emph{New method for high-accuracy
  determination of the fine-structure constant based on quantized {H}all
  resistance}, \href{https://doi.org/10.1103/PhysRevLett.45.494}{\emph{Phys.
  Rev. Lett.} {\bfseries 45} (1980) 494}.

\bibitem{Qi2011:RMP}
X.-L.~Qi and S.-C.~Zhang, \emph{Topological insulators and superconductors},
  \href{https://doi.org/10.1103/RevModPhys.83.1057}{\emph{Rev. Mod. Phys.}
  {\bfseries 83} (2011) 1057}
  [\href{https://arxiv.org/abs/1008.2026}{{\ttfamily 1008.2026}}].

\bibitem{Hasan2010:RMP}
M.Z.~Hasan and C.L.~Kane, \emph{Colloquium: Topological insulators},
  \href{https://doi.org/10.1103/RevModPhys.82.3045}{\emph{Rev. Mod. Phys.}
  {\bfseries 82} (2010) 3045}
  [\href{https://arxiv.org/abs/1002.3895}{{\ttfamily 1002.3895}}].

\bibitem{Wannier1972:PRB}
G.H.~Wannier, \emph{Theorem on the magnetoconductivity of metals},
  \href{https://doi.org/10.1103/PhysRevB.5.3836}{\emph{Phys. Rev. B} {\bfseries
  5} (1972) 3836}.

\bibitem{Kim2013:PRL}
H.-J.~Kim, K.-S.~Kim, J.-F.~Wang, M.~Sasaki, N.~Satoh, A.~Ohnishi et~al.,
  \emph{Dirac versus {W}eyl fermions in topological insulators:
  {A}dler-{B}ell-{J}ackiw anomaly in transport phenomena},
  \href{https://doi.org/10.1103/PhysRevLett.111.246603}{\emph{Phys. Rev. Lett.}
  {\bfseries 111} (2013) 246603}
  [\href{https://arxiv.org/abs/1307.6990}{{\ttfamily 1307.6990}}].

\bibitem{Xiong2015:Sci}
J.~Xiong, S.K.~Kushwaha, T.~Liang, J.W.~Krizan, M.~Hirschberger, W.~Wang
  et~al., \emph{Evidence for the chiral anomaly in the {D}irac semimetal
  {N}a$_3${B}i}, \href{https://doi.org/10.1126/science.aac6089}{\emph{Science}
  {\bfseries 350} (2015) 413}.

\bibitem{Li2016:NC}
H.~Li, H.~He, H.-Z.~Lu, H.~Zhang, H.~Liu, R.~Ma et~al., \emph{Negative
  magnetoresistance in {D}irac semimetal cd$_3$as$_2$},
  \href{https://doi.org/10.1038/ncomms10301}{\emph{Nature communications}
  {\bfseries 7} (2016) 1} [\href{https://arxiv.org/abs/1507.06470}{{\ttfamily
  1507.06470}}].

\bibitem{Zhang2016:NC}
S.-Y.X.~Cheng-Long~Zhang et~al., \emph{Signatures of the
  {A}dler-{B}ell-{J}ackiw chiral anomaly in a {W}eyl fermion semimetal},
  \href{https://doi.org/10.1038/ncomms10735}{\emph{Nature communications}
  {\bfseries 7} (2016) 1} [\href{https://arxiv.org/abs/1601.04208}{{\ttfamily
  1601.04208}}].

\bibitem{Zhao2016:SR}
B.~Zhao, P.~Cheng, H.~Pan, S.~Zhang, B.~Wang, G.~Wang et~al., \emph{Weak
  antilocalization in cd$_3$as$_2$ thin films},
  \href{https://doi.org/10.1038/srep22377}{\emph{Scientific Reports} {\bfseries
  6} (2016) 22377} [\href{https://arxiv.org/abs/1601.05536}{{\ttfamily
  1601.05536}}].

\bibitem{DTSon2013:PRB}
D.T.~Son and B.Z.~Spivak, \emph{Chiral anomaly and classical negative
  magnetoresistance of {W}eyl metals},
  \href{https://doi.org/10.1103/PhysRevB.88.104412}{\emph{Phys. Rev. B}
  {\bfseries 88} (2013) 104412}
  [\href{https://arxiv.org/abs/1206.1627}{{\ttfamily 1206.1627}}].

\bibitem{Qi2013:CRP}
P.~Hosur and X.~Qi, \emph{Recent developments in transport phenomena in {W}eyl
  semimetals}, \href{https://doi.org/10.1016/j.crhy.2013.10.010}{\emph{Comptes
  Rendus Physique} {\bfseries 14} (2013) 857}
  [\href{https://arxiv.org/abs/1309.4464}{{\ttfamily 1309.4464}}].

\bibitem{Burkov2014:PRL}
A.A.~Burkov, \emph{Chiral anomaly and diffusive magnetotransport in {W}eyl
  metals}, \href{https://doi.org/10.1103/PhysRevLett.113.247203}{\emph{Phys.
  Rev. Lett.} {\bfseries 113} (2014) 247203}
  [\href{https://arxiv.org/abs/1409.0013}{{\ttfamily 1409.0013}}].

\bibitem{Burkov2015:PRB}
A.A.~Burkov, \emph{Negative longitudinal magnetoresistance in {D}irac and
  {W}eyl metals}, \href{https://doi.org/10.1103/PhysRevB.91.245157}{\emph{Phys.
  Rev. B} {\bfseries 91} (2015) 245157}
  [\href{https://arxiv.org/abs/1505.01849}{{\ttfamily 1505.01849}}].

\bibitem{Lu2017:FP}
H.-Z.~Lu and S.-Q.~Shen, \emph{Quantum transport in topological semimetals
  under magnetic fields},
  \href{https://doi.org/10.1007/s11467-016-0609-y}{\emph{Frontiers of Physics}
  {\bfseries 12} (2017) 127201}
  [\href{https://arxiv.org/abs/1609.01029}{{\ttfamily 1609.01029}}].

\bibitem{Jimenez2014:PRD}
A.~Jimenez-Alba, K.~Landsteiner and L.~Melgar, \emph{Anomalous magnetoresponse
  and the {S}t{\"u}ckelberg axion in holography},
  \href{https://doi.org/10.1103/PhysRevD.90.126004}{\emph{Phys. Rev. D}
  {\bfseries 90} (2014) 126004}
  [\href{https://arxiv.org/abs/1407.8162}{{\ttfamily 1407.8162}}].

\bibitem{Jimenez2015:JHEP}
A.~Jimenez-Alba, K.~Landsteiner, Y.~Liu and Y.-W.~Sun, \emph{Anomalous
  magnetoconductivity and relaxation times in holography},
  \href{https://doi.org/10.1007/JHEP07(2015)117}{\emph{Journal of High Energy
  Physics} {\bfseries 2015} (2015) 117}
  [\href{https://arxiv.org/abs/1504.06566}{{\ttfamily 1504.06566}}].

\bibitem{Landsteiner2015:JHEP}
K.~Landsteiner, Y.~Liu and Y.-W.~Sun, \emph{Negative magnetoresistivity in
  chiral fluids and holography},
  \href{https://doi.org/10.1007/JHEP03(2015)127}{\emph{Journal of High Energy
  Physics} {\bfseries 2015} (2015) 127}
  [\href{https://arxiv.org/abs/1410.6399}{{\ttfamily 1410.6399}}].

\bibitem{Sun2016:JHEP}
Y.-W.~Sun and Q.~Yang, \emph{Negative magnetoresistivity in holography},
  \href{https://doi.org/10.1007/JHEP09(2016)122}{\emph{Journal of High Energy
  Physics} {\bfseries 2016} (2016) 122}
  [\href{https://arxiv.org/abs/1603.02624}{{\ttfamily 1603.02624}}].

\bibitem{Wen2006:RMP}
P.A.~Lee, N.~Nagaosa and X.-G.~Wen, \emph{Doping a {M}ott insulator: Physics of
  high-temperature superconductivity},
  \href{https://doi.org/10.1103/RevModPhys.78.17}{\emph{Rev. Mod. Phys.}
  {\bfseries 78} (2006) 17}
  [\href{https://arxiv.org/abs/cond-mat/0410445}{{\ttfamily
  cond-mat/0410445}}].

\bibitem{Edalati2011:PRL}
M.~Edalati, R.G.~Leigh and P.W.~Phillips, \emph{Dynamically generated {M}ott
  gap from holography},
  \href{https://doi.org/10.1103/PhysRevLett.106.091602}{\emph{Phys. Rev. Lett.}
  {\bfseries 106} (2011) 091602}
  [\href{https://arxiv.org/abs/1010.3238}{{\ttfamily 1010.3238}}].

\bibitem{Edalati2011:PRD}
M.~Edalati, R.G.~Leigh, K.W.~Lo and P.W.~Phillips, \emph{Dynamical gap and
  cupratelike physics from holography},
  \href{https://doi.org/10.1103/PhysRevD.83.046012}{\emph{Phys. Rev. D}
  {\bfseries 83} (2011) 046012}
  [\href{https://arxiv.org/abs/1012.3751}{{\ttfamily 1012.3751}}].

\bibitem{Wu2012:JHEP}
J.-P.~Wu and H.-B.~Zeng, \emph{Dynamic gap from holographic fermions in charged
  dilaton black branes},
  \href{https://doi.org/10.1007/JHEP04(2015)137}{\emph{Journal of High Energy
  Physics} {\bfseries 2012} (2012) 68}
  [\href{https://arxiv.org/abs/1411.5627}{{\ttfamily 1411.5627}}].

\bibitem{Wu2014:JHEP}
Y.~Ling, P.~Liu, C.~Niu, J.-P.~Wu and Z.-Y.~Xian, \emph{Holographic fermionic
  system with dipole coupling on {Q}-lattice},
  \href{https://doi.org/10.1007/JHEP12(2014)149}{\emph{Journal of High Energy
  Physics} {\bfseries 2014} (2014) 149}
  [\href{https://arxiv.org/abs/1410.7323}{{\ttfamily 1410.7323}}].

\bibitem{Wu2015:JHEP}
Y.~Ling, P.~Liu, C.~Niu and J.-P.~Wu, \emph{Building a doped mott system by
  holography}, \href{https://doi.org/10.1103/PhysRevD.92.086003}{\emph{Phys.
  Rev. D} {\bfseries 92} (2015) 086003}
  [\href{https://arxiv.org/abs/1507.02514}{{\ttfamily 1507.02514}}].

\bibitem{Wu2016:JHEP}
Y.~Ling, P.~Liu and J.-P.~Wu, \emph{A novel insulator by holographic
  {Q}-lattices}, \href{https://doi.org/10.1007/jhep02(2016)075}{\emph{Journal
  of High Energy Physics} {\bfseries 2016} (2016) 75}.

\bibitem{Fujita2015:JHEP}
M.~Fujita, S.M.~Harrison, A.~Karch, R.~Meyer and N.M.~Paquette, \emph{Towards a
  holographic bose-hubbard model},
  \href{https://doi.org/10.1007/JHEP04(2015)068}{\emph{Journal of High Energy
  Physics} {\bfseries 2015} (2015) 68}
  [\href{https://arxiv.org/abs/1411.7899}{{\ttfamily 1411.7899}}].

\bibitem{Kiritsis2015:JHEP}
E.~Kiritsis and J.~Ren, \emph{On holographic insulators and supersolids},
  \href{https://doi.org/10.1007/JHEP09(2015)168}{\emph{Journal of High Energy
  Physics} {\bfseries 2015} (2015) 168}
  [\href{https://arxiv.org/abs/1503.03481}{{\ttfamily 1503.03481}}].

\bibitem{Baggioli2016:JHEP}
M.~Baggioli and O.~Pujolas, \emph{On effective holographic {M}ott insulators},
  \href{https://doi.org/10.1007/JHEP12(2016)107}{\emph{Journal of High Energy
  Physics} {\bfseries 2016} (2016) 107}
  [\href{https://arxiv.org/abs/1604.08915}{{\ttfamily 1604.08915}}].

\bibitem{Cremonini2017:JHEP}
S.~Cremonini, A.~Hoover and L.~Li, \emph{Backreacted {DBI} magnetotransport
  with momentum dissipation},
  \href{https://doi.org/10.1007/JHEP10(2017)133}{\emph{Journal of High Energy
  Physics} {\bfseries 2017} (2017) 133}
  [\href{https://arxiv.org/abs/1707.01505}{{\ttfamily 1707.01505}}].

\bibitem{Vegh2013}
D.~Vegh, \emph{Holography without translational symmetry},
  \href{https://arxiv.org/abs/1301.0537}{{\ttfamily 1301.0537}}.

\bibitem{Cai2015:PRD}
R.-G.~Cai, Y.-P.~Hu, Q.-Y.~Pan and Y.-L.~Zhang, \emph{Thermodynamics of black
  hooles in massive gravity},
  \href{https://doi.org/10.1103/PhysRevD.91.024032}{\emph{Phys. Rev. D}
  {\bfseries 91} (2015) 024032}
  [\href{https://arxiv.org/abs/1409.2369}{{\ttfamily 1409.2369}}].

\bibitem{Donos2014:JHEP}
A.~Donos and J.P.~Gauntlett, \emph{Novel metals and insulators from
  holography}, \href{https://doi.org/10.1007/JHEP06(2014)007}{\emph{Journal of
  High Energy Physics} {\bfseries 2014} (2014) 7}
  [\href{https://arxiv.org/abs/1401.5077}{{\ttfamily 1401.5077}}].

\bibitem{Blake2014:PRL}
M.~Blake and A.~Donos, \emph{Quantum critical transport and the {H}all angle in
  holographic models},
  \href{https://doi.org/10.1103/PhysRevLett.114.021601}{\emph{Phys. Rev. Lett.}
  {\bfseries 114} (2015) 021601}
  [\href{https://arxiv.org/abs/1406.1659}{{\ttfamily 1406.1659}}].

\bibitem{Hartnoll2007:PRD}
S.A.~Hartnoll and P.K.~Kovtun, \emph{{H}all conductivity from dyonic black
  holes}, \href{https://doi.org/10.1103/PhysRevD.76.066001}{\emph{Phys. Rev. D}
  {\bfseries 76} (2007) 066001}
  [\href{https://arxiv.org/abs/0704.1160}{{\ttfamily 0704.1160}}].

\bibitem{Guo2017:PRD}
X.~Guo, P.~Wang and H.~Yang, \emph{Membrane paradigm and holographic {DC}
  conductivity for nonlinear electrodynamics},
  \href{https://doi.org/10.1103/PhysRevD.98.026021}{\emph{Phys. Rev. D}
  {\bfseries 98} (2018) 026021}
  [\href{https://arxiv.org/abs/1711.03298}{{\ttfamily 1711.03298}}].

\bibitem{Davison2015:JHEP}
R.A.~Davison and B.~Gout{\'e}raux, \emph{Dissecting holographic
  conductivities}, \href{https://doi.org/10.1007/JHEP09(2015)090}{\emph{Journal
  of High Energy Physics} {\bfseries 2015} (2015) 90}
  [\href{https://arxiv.org/abs/1505.05092}{{\ttfamily 1505.05092}}].

\bibitem{Peng2018:EPJC}
P.~Wang, H.~Wu and H.~Yang, \emph{Holographic {DC} conductivity for backreacted
  nonlinear electrodynamics with momentum dissipation},
  \href{https://doi.org/10.1140/epjc/s10052-018-6503-8}{\emph{Eur. Phys. J. C}
  {\bfseries 79} (2019) 6} [\href{https://arxiv.org/abs/1805.07913}{{\ttfamily
  1805.07913}}].

\bibitem{Zhang1989:PRL}
S.C.~Zhang, T.H.~Hansson and S.~Kivelson, \emph{Effective-field-theory model
  for the fractional quantum {H}all effect},
  \href{https://doi.org/10.1103/PhysRevLett.62.82}{\emph{Phys. Rev. Lett.}
  {\bfseries 62} (1989) 82}.

\bibitem{Fradkin1991:PRB}
A.~Lopez and E.~Fradkin, \emph{Fractional quantum {H}all effect and
  {C}hern-{S}imons gauge theories},
  \href{https://doi.org/10.1103/PhysRevB.44.5246}{\emph{Phys. Rev. B}
  {\bfseries 44} (1991) 5246}.

\bibitem{Zhang1992:IJMPB}
S.C.~Zhang, \emph{The {C}hern-{S}imons-{L}andau-{G}inzburg theory of the
  fractional quantum {H}all effect},
  \href{https://doi.org/10.1142/S0217979292000037}{\emph{International Journal
  of Modern Physics B} {\bfseries 6} (1992) 25}.

\bibitem{DavidTong:QHE}
D.~Tong, \emph{Lectures on the quantum hall effect},
  \href{https://arxiv.org/abs/1606.06687}{{\ttfamily 1606.06687}}.

\bibitem{Born1933:Nat}
M.~Born, \emph{Modified field equations with a finite radius of the electron},
  \href{https://doi.org/10.1038/132282a0}{\emph{Nature} {\bfseries 132} (1933)
  282}.

\bibitem{Born1934:PRSLA}
M.~Born, \emph{On the quantum theory of the electromagnetic field},
  \href{https://doi.org/10.1098/rspa.1934.0010}{\emph{Proceedings of the Royal
  Society of London. Series A, Containing Papers of a Mathematical and Physical
  Character} {\bfseries 143} (1934) 410}.

\bibitem{zwiebach2004:CUP}
B.~Zwiebach, \emph{A First Course in String Theory}, Cambridge University Press
  (2009).

\end{thebibliography}\endgroup

\end{document}